\documentclass{aa}  
\usepackage{graphicx}
\usepackage{longtable}
\usepackage{subcaption}
\usepackage{lscape}
\usepackage{eqnarray,amsmath}
\usepackage{booktabs} 
\usepackage{makecell}   
\usepackage{amsmath}   
\usepackage{float}
\usepackage{afterpage}
\usepackage{txfonts}
\usepackage{xcolor}
\usepackage[colorlinks=true, citecolor=blue, linkcolor=blue, urlcolor=blue]{hyperref}
\begin{document} 
\title{Three-dimensional GRMHD simulations of 
jet formation and propagation in self-gravitating collapsing stars}
\titlerunning{Three-dimensional GRMHD simulations of jet formation and propagation}

   \author{Piotr Płonka\inst{1}\fnmsep\thanks{pplonka@cft.edu.pl}
          \and
          Agnieszka Janiuk\inst{1}\fnmsep\thanks{agnes@cft.edu.pl}
          }
            
              \institute{Center for Theoretical Physics, Polish Academy of Sciences, Al. Lotnik\'ow 32/46, 02-668 Warsaw, Poland\\
                                }


 
  \abstract
   {Discovered almost sixty years ago, gamma-ray bursts are the most powerful explosions in the Universe. Long gamma-ray bursts are associated with the collapse of rapidly rotating massive stars, which conclude their lives as stellar-mass black holes. During this process, the initial mass of the black hole is several times smaller than the remaining mass of the progenitor. Taking into account self-gravity of the star may significantly modify the process of jet formation. This potentially affects the prompt emission.}
   {We investigate collapsar models with and without self-gravity under identical initial conditions to directly compare the effects of self-gravity on jet properties, such as opening angle, jet power, terminal Lorentz factor, and its variability.}
   {We compute a suite of time-dependent, three-dimensional GRMHD simulations of collapsars in evolving spacetime. We update the Kerr metric components due to the growth of the black hole mass and changes its angular momentum. The self-gravity is considered via perturbative terms.}
   {We present for the first time the process of jet formation in self-gravitating collapsars. We find that self-gravity leads to temporary jet quenching, which can explain some features in the gamma-ray burst prompt emission. We find no substantial difference in jet launching times between models with and without self-gravity. We observe that in the absence of self-gravity, the jet can extract more rotational energy from the black hole, while self-gravitating models produce narrower jet opening angles. We show that under certain conditions, self-gravity can interrupt the jet formation process, resulting in a failed burst.}
   {Our computations show that self-gravity significantly modifies the process of jet propagation, resulting in notably different jet properties. We show that the timescales, variability, and opening angle of jet depend on whether self-gravity is included or not. We argue that self-gravity can potentially explain certain prompt emission properties due to the jet quenching.}

\keywords{Gamma-ray burst: general -- Stars: jets -- Accretion, accretion disks -- Black hole physics -- Magnetohydrodynamics (MHD) -- Stars: massive}
   \maketitle
%

\section{Introduction}
The evolution of a star is primarily determined by its mass, metallicity, and rotation rate \citep{1997A&A...321..465M,Janaka2007}. Massive stars, when nuclear fusion stops producing energy, collapse. If their mass exceeds about \mbox{$\sim25\,M_{\odot}$}, they conclude their lives as stellar black holes \citep{2003ApJ...591..288H}. Gamma-ray bursts (GRBs) are classified based on their duration into two classes: short and long \citep{1993ApJ...413L.101K}. Long gamma-ray bursts are typically associated with Type Ib and Type Ic supernovae, which implies that their progenitors have lost their outer layers \citep{Woosley2006}. Nevertheless, new detections of long-duration GRBs without SN association have questioned the simple duration-defined dichotomy, and motivated multi-parameter class definitions as proposed in \cite{2009ApJ...703.1696Z}. 
The collapsar model is currently the most widely acknowledged explanation for the origin of long GRBs \citep{Woosley1993,1999ApJ...524..262M}. The typical energy release in such an event is between $10^{51}$ and $10^{54}$ erg, with durations ranging from about 2 to 1000 seconds \citep{Piran2004,Kumran2015}.\\
\indent Variability demonstrated by the prompt phase is one of the key observables used to constrain theoretical models of GRB progenitors. Interaction of shocks emitted with slightly different Lorentz factors within the GRB jet was proposed by \cite{1998MNRAS.296..275D}. In addition, the collisionless shocks may leave an imprint in the afterglow emission \citep{2008ApJ...673L..39S}. The prompt phase has a rapid onset, and then the resulting lightcurve can vary over a range of timescales, showing sometimes multiple distinct, spiky episodes spread over the burst duration \citep{Kumran2015}. 
Observations of afterglow lightcurves suggest a long-lived central engine, a wide distribution of Lorentz factors in the jets, and possibly the deceleration of a Poynting flux-dominated flow \citep{2006ApJ...642..354Z}.\\
\indent In the collapsar model, the jet drills through the progenitor star, and the jet-star interaction may produce structures in the lightcurve. This imprint could filter the variability arising primarily from the central engine \citep{Morsony_2010}. After breakout, the observed luminosity variations record the central engine variability carried by internal dissipation in the outflow due to shocks or magnetic reconnections \citep{Zhang_2011,2016MNRAS.456.1739B}, which are both capable of reproducing the highly variable prompt emission seen in long GRBs \citep{1997ApJ...490...92K}. The collapsar scenario also accommodates longer “quiescent” gaps and late X-ray flares \citep{2011MNRAS.410.1064M} by invoking intermittent engine activity, when the jet can choke. 
The spectral and timing properties of prompt and X-ray flares point towards dissipation from renewed variable engine activity \citep{2006ApJ...642..354Z}.
The alternative to a hyperaccreting black hole in the collapsing star is a magnetar model where different kinds of instabilities operate \citep{2009MNRAS.396.2038B}.
However, the observed properties of some GRBs are in contrast with this scenario and imply a strong evolution of Lorentz factor and photospheric radius \citep[e.g.]{2013MNRAS.433.2739I} that can be reconciled with an evolving black hole central engine.

Numerical simulations of collapsars often simplify the hyperaccretion phase and process of jet ejection by assuming a fixed black hole spin, its constant mass, and uniform flow magnetisation. If those parameters evolve, the recovery of prompt and early afterglow phases phenomenology is more appealing. Even modest changes of the Kerr parameter, or magnetic flux, can map into jet luminosity evolution \citep{Tchekhovskoy2011}. 
It has been shown that MAD/SANE transitions and flux eruption episodes can intermittently choke accretion, giving rise to multiple pulses and gaps due to engine “on/off” behaviour, which match Swift-era flare phenomenology much better than a fixed central engine schematic \citep{2016MNRAS.461.1045L}. Recently, axisymmetric GRMHD models in dynamical spacetime have pointed towards characteristic spin changes on long timescales, relevant for the overall prompt phase duration \citep{PhysRevD.109.043051}.

\indent In our previous two-dimensional study, we investigated the black hole mass and spin evolution, and we also incorporated effects of self-gravity during the collapse \citep{JaniukSG}. We demonstrated that for self-gravitating collapsars, the central engine evolves faster and the accretion rate exhibits more rapid variability than in the models without self-gravity. We showed that if the infalling stellar envelope mass rivals or exceeds the seed BH mass, the disk–BH system is strongly perturbed and self-gravity enhances shocks and large-amplitude $\dot M$ oscillations. In those simulations, we were not able to produce successful jets.
In this study, we present three-dimensional simulations with jets emerging from self-gravitating stars. 

\indent The article is organized as follows. In Sect.~\ref{sec:Initial_conditions}, we define the initial conditions of our models. 
Then, we give the equations used to implement self-gravity in the code in Sect.~\ref{sec:Self-gravity_implementation}, and finally, we give the evolution equations and methods used by the GRMHD code in Sect.~\ref{sec:Numerical_implementation}. In Sect.~\ref{sec:Jet_energetics_and_properties}, we present our results about the effects of self-gravity on the jet energetics and properties. In Sect.~\ref{sec:Jet_opening_angle}, we focus on the effects of self-gravity on the opening angle of relativistic jets. In Sect.~\ref{sec:Time-dependent evolution}, we concern the time-dependent evolution of the central black hole. In Sect.~\ref{sec:Accretion_flow_and_MAD_state}, we discuss accretion flows and the stability of the MAD state. In Sect.~\ref{sec:Discussion}, we discuss our results in the context of GRB prompt emission, and in Sect.~\ref{sec:Conclusions}, we summarize our conclusions.

\section{Simulation setup}
\subsection{Initial conditions}
\label{sec:Initial_conditions}
We use the initial conditions as described in detail in
\cite{Janiuk2018,Dominika2021,JaniukSG}, for the
quasi-spherical accretion. Our collapsar models assume that the entire configuration has had enough time to achieve equilibrium. Therefore, we start with the Bondi solution \citep{1952MNRAS.112..195B}, endowed with a small angular momentum. 
The total mass available for accretion is defined at the start, and no additional matter is added later on (except numerical flooring). 
The equation of state is polytropic, given by \mbox{$P = K \rho^\gamma$}, where $P$ is the pressure,  $\rho$ is the density, and $K$ refers to the specific entropy. The relation is parameterized by the polytropic index \mbox{$\gamma={4}/{3}$}. The sonic radius $r_s$ is set at the distance of \mbox{$80\,r_g$} from the central black hole, where $r_g$ is 
the gravitational radius. Below this point, matter falls supersonically towards the black hole. The radial velocity $u^{r}_{s}$ at critical point is evaluated from:
\begin{equation}
(u^{r}_{s})^{2} = \frac{1}{2r_{s}}.
\end{equation}
The profile of radial velocity \(u^{r}\) as a function of $r$ is obtained by numerically solving the relativistic Bernoulli equation \citep{1972Ap&SS..15..153M,shapiro1983}, 
given by:
\begin{equation}
\left( 1 + \frac{\gamma}{\gamma - 1} \frac{P}{\rho} \right)^2 \left( 1 - \frac{2}{r} + (u^r)^2 \right) = \text{constant}.
\end{equation}
The density distribution is evaluated based on the radial velocity, and scaled by mass accretion rate $\dot{M}$, which is adjusted to satisfy the specified target for the total mass of the evolved stellar core $M_\star$ in the simulation domain, and is expressed by:
\begin{equation}
\displaystyle \rho = \frac{\dot{M}}{4\pi r^2 u^r}.
\end{equation}
The specific entropy is calculated at the sonic radius from the following formula:
\begin{equation}
\begin{array}{ll}
\displaystyle
K = \left(u^{r} 4\pi r^{2}\,\frac{\displaystyle c_{s}^{\frac{2}{\gamma - 1}}}{\displaystyle \gamma^{\frac{1}{\gamma - 1}}\, \dot{M}}
\right)^{\gamma - 1},&\displaystyle c_s^{2} =\displaystyle \gamma\,\displaystyle\frac{\displaystyle P}{\displaystyle \rho},
\end{array}
\end{equation}
where $c_s$ is the speed of sound.\\
\indent The angular momentum of the stellar core is scaled 
with $l_{\text{isco}}$, representing the angular momentum of a particle orbiting the black hole at the innermost stable circular orbit \citep{Sukova2015,2017MNRAS.472.4327S,2019MNRAS.487..755P}. In the Kerr metric, the radius of the ISCO, expressed in units of $r_g$, is given by:
\begin{equation}
r_{\mathrm{isco}} = 3 + Z_2 \mp \sqrt{(3 - Z_1)(3 + Z_1 + 2Z_2)},
\end{equation}

\begin{equation}
Z_1 \equiv 1 + (1 - a^2)^{1/3} \left[ (1 + a)^{1/3} + (1 - a)^{1/3} \right],
\end{equation}

\begin{equation}
Z_2 \equiv \sqrt{3a^2 + Z_1^2},
\end{equation}
where $a$ is the spin of the black hole. 
For a Kerr black hole, \(l_{\text{isco}}\) and auxiliary energy $\epsilon_{\text{isco}}$ depend only on the mass and the spin of the black hole \citep{1972ApJ...178..347B}, and are given by:
\begin{equation}
    l_{\text{isco}} = \frac{r_{\text{isco}}^{1/2} - 2a / r_{\text{isco}} + a^2 / r_{\text{isco}}^{3/2}}
    {\sqrt{1 - 3 / r_{\text{isco}} + 2a / r_{\text{isco}}^{3/2}}},
\end{equation}
\begin{equation}
\epsilon_{\text{isco}} = \frac{1 - 2 / r_{\text{isco}} + a / r_{\text{isco}}^{3/2}} 
{\sqrt{1 - 3 / r_{\text{isco}} + 2a / r_{\text{isco}}^{3/2}}}.
\end{equation}
\noindent The distribution of the angular velocity $u^{\phi}$ is given by: 
\begin{equation}
u^{\phi}(r, \theta) = S\,\big(-g^{t\phi} \, \epsilon_{\text{isco}} + g^{\phi\phi} \, l_{\text{isco}} \big) \sin^{2}\theta,
\label{eq:u_phi}
\end{equation}
where the $g^{t\phi}$ and $g^{\phi\phi}$ are the metric tensor components 
in the Boyer–Lindquist coordinates given by \mbox{$g^{t\phi} = -2ar/\Sigma\Delta$}, \mbox{$g^{\phi\phi}=(\Delta-a^2\sin^2\theta)/(\Sigma\Delta\sin^2\theta)$}, \mbox{$\Sigma = r^2 + a^2 \cos^2 \theta$}, and $\Delta = r^2 - 2r + a^2$. 
The dimensionless parameter $S$ is used to scale  
the angular momentum with respect to ISCO. For \mbox{$S=0$}, the model describes purely spherical accretion. 
The distribution of the angular momentum is additionally multiplied by $\sin^2 \theta$
to ensure the maximum rotational velocity at the equatorial plane \citep{2017MNRAS.472.4327S}, where an accretion disk should form \citep{1999ApJ...524..262M}.\\
\indent The collapsar models
producing relativistic jets should be
highly magnetized \citep{10.1093/mnras/stu2229,Gottlieb2022}. 
To initialize the magnetic field, we use the plasma  parameter $\beta$, defined as the ratio of the gas pressure, \mbox{$p = (\gamma - 1)u$}, where $u$ is the specific internal energy,
to the magnetic pressure, \mbox{$p_{\mathrm{mag}} = \tfrac{1}{2} b^{\mu} b_{\mu}$}, where $b^{\mu}$ is the contravariant component of the magnetic four-vector. Our models are scaled to 
the value of  
$\beta_0$ at \mbox{$2\,{r_g}$}. We adopted  
a uniform magnetic field given by the magnetic potential $\mathbf{A}_\varphi$, defined as:
\begin{equation}
\mathbf{A}_\varphi \propto \left( 0,\ 0,\ \frac{1}{2}r\sin\theta \right).
\end{equation}

The initial mass of the black hole 
is assumed as \mbox{$M_{0}=3\,M_{\odot}$}, whereas the mass of the stellar core is \mbox{$M_{\star}=25\,M_{\odot}$}. The initial spin of the black holes is chosen to be \mbox{$a_0=0.8$}. The angular momentum is scaled 
with \mbox{$S=2$} in all models, similar to \citet{Murguia-Berthier_2020}. The magnetic field is scaled with \mbox{$\beta_0 = 1$} or \mbox{$\beta_0 = 10$}, corresponding approximately to 
field strengths ranging from \mbox{$B\sim10^{8}\,\mathrm{G}$} in the outer region to \mbox{$B\sim10^{15}\,\mathrm{G}$} in the innermost core.
We motivate this choice by the magnetic fields of 
\mbox{$B\sim10^{15}\,\mathrm{G}$} observed in magnetars \citep{2015RPPh...78k6901T,2017ARA&A..55..261K}. 

In addition,
to break the azimuthal symmetry, we introduce the initial random perturbation in the internal energy. 
The amplitude of the perturbation is set to \mbox{$5\,\%$}, as in \citet{2018MNRAS.479.2534M}. All models are three-dimensional. The resolution of runs, and their initial parameters 
are listed in Table \ref{tab:models}.

\begin{table*}[t!]
\centering
\begin{tabular}{|l|c|c|c|c|c|c|c|c|c|c|c|}
\specialrule{0.4pt}{0pt}{2pt}  
\specialrule{0.4pt}{0.3pt}{0pt} 
\noalign{\vskip 0.5mm}
\texttt{{Model}} & \makecell{$\beta_0$ \\ \vphantom{x}} & \makecell{$M_{\mathrm{0}}$ \\ $(M_{\odot})$} & \makecell{$M_{\star}$ \\ $(M_{\odot})$} &  \makecell{$a_{0}$ \\ \vphantom{x}} & \makecell{$S$ \\ \vphantom{x}} & \makecell{Magnetic \\ Field} & \makecell{Perturbation \\ $(\%)$ } & \makecell{Self \\ Gravity}  & \makecell{Resolution \\$( N_r \times N_{\theta} \times N_{\phi})$} & \makecell{$t_{{f}}$ \\ $(t_{g})$} \\
\midrule
\texttt{{Model-1-SG}} & 1  & 3  & 25  & 0.8 & 2  & vertical & 5  & yes & $384 \times 192 \times 128$ & $5\times10^4$\\[0.5mm]
\texttt{{Model-2-SG}} & 10  & 3  & 25  & 0.8 & 2  & vertical  & 5 & yes  & $256 \times 128 \times 64$& $5\times10^4$\\ [0.5mm]
\hline 
\noalign{\vskip 0.5mm}
\texttt{{Model-1-NSG}} & 1  & 3  & 25  & 0.8 & 2  & vertical & 5 & no  & $384 \times 192 \times 128$ & $5\times10^4$\\[0.5mm]
\texttt{{Model-2-NSG}} & 10  & 3  & 25  & 0.8 & 2  & vertical  & 5 & no  & $256 \times 128 \times 64$& $5\times10^4$\\
\specialrule{0.4pt}{0pt}{0pt}
\end{tabular}
\caption{Models considered in this study for the initial black hole of \mbox{$M_{0}=3\,M_\odot$} and the initial stellar core mass of \mbox{$M_{\star}=25\,M_\odot$}.}

\label{tab:models}
\end{table*}

\subsection{Self-gravity implementation}
\label{sec:Self-gravity_implementation}
Typical GRMHD models of collapsars assume that the central black hole has already formed, and it is surrounded by the accretion disk fuelled via matter fallback from the massive stellar envelope \citep{Mizuno_2004}. For the stripped-envelope star, its mass is still several times larger than the initial black hole mass \citep{2003ApJ...591..288H,WoosleyHeger2006,2007ARA&A..45..177C}. Therefore, self-gravity cannot be ignored in physically consistent simulations. Although a complete numerical solution of the full Einstein equations in a three-dimensional collapsar scenario is theoretically possible, its computational complexity is very high. In this work, we adopt an alternative technique to model self-gravity, allowing for an efficient run of a set of three-dimensional models. In this way, we aim to investigate the general impact of self-gravity on the magnetized collapsing stars.\\
\indent Evolution of the mass and spin of the black hole was considered, without self-gravitating force, in \citet{Janiuk2018, Dominika2021}. The implementation presented in \citet{JaniukSG}, took into account self-gravity as perturbative terms to Kerr metric to analyze simultaneously black hole mass and spin evolution, as well as self-gravity effects. Our approach is motivated by the Teukolsky equation \citep{1972PhRvL..29.1114T}. The global vacuum solution is formulated using the CCK formalism \citep{Chrzanowski1975, Cohen1974, Wald1978}, enabling us to make the reconstruction of the metric perturbations, following the approach of \cite{2017CQGra..34l4003V} who proved that the perturbation caused by a particle in any bound orbit around a Kerr black hole affects only two parameters, $a$ and $M$, measured above the particle’s orbit. Notably, the perturbation disappears inside the orbit.\\
\indent The modifications of metric mass coefficients in our code are divided into two groups. The first group is responsible for the changes in the central black hole mass \(\Delta M\), due to the mass-energy flux crossing the black hole horizon (primarily mass, but also other forms of energy contained in the radial energy flux \(T^{r}_{t}\)). This flux is integrated over the black hole horizon $r_h$ (note that the size of the event horizon depends on time) at each time step of the simulation using the following formula:
\begin{equation}
\Delta M(t) = \int_{0}^{t} \int_{0}^{2\pi} \int_{0}^{\pi} \sqrt{-g} \, T^{r}_{t} \, d\theta \, d\phi \, dt.
\end{equation}
\noindent The second group comes from the perturbative term \(\delta M\), which is responsible for the self-gravity and depends on both time and radius. We calculate this perturbation by integrating the energy density \(T^{t}_{t}\) 
over the spherical volume enclosed between the event horizon \(r_h\) and the given radius \(r\) as follows:
\begin{equation}
\delta M(t,r) = \int_{r_{\text{h}}}^{r} \int_{0}^{2\pi} \int_{0}^{\pi} \sqrt{-g} \, T^{t}_{t} \, d\theta \, d\phi \, dr.
\end{equation}
\noindent The evolution of the mass metric coefficient is finally given by the following three terms:
\begin{equation}
M(t,r) = M_0 + \Delta M(t) + \delta M(t,r)
\end{equation}
\indent Similarly, we introduce the dynamical evolution of the black hole's angular momentum \(J(t)\). 
The initial angular momentum is \(J_0\), and its evolution $\Delta J(t)$ is governed by the angular momentum flux \(T^r_\phi\) crossing the black hole horizon, given by: 
\begin{equation}
\label{eq:spin_angular_momentum}
\Delta J(t) = \int_{0}^{t} \int_{0}^{2\pi} \int_{0}^{\pi} \sqrt{-g} \, T^r_{\phi} \, d\theta \, d\phi \, dt.
\end{equation}
The second component is the perturbative term \(\delta J(t,r)\), which accounts for the total angular momentum density $T^t_{\phi}$ contained in the volume enclosed between the event horizon \(r_h\) and the given radius \(r\). The term is computed as:
\begin{equation}
\delta J(t,r) = \int_{r_{\text{h}}}^{r} \int_{0}^{2\pi} \int_{0}^{\pi} \sqrt{-g} \, T^t_{\phi} \, d\theta \, d\phi \, dr.
\end{equation}
\noindent The dynamical evolution of the angular momentum is finally given by:
\begin{equation}
J(t,r) = J_0 + \Delta J(t) + \delta J(t,r).
\end{equation}
\indent To sum up, we have two dynamical metric components in Kerr--Schild coordinates, depending on time and distance from the black hole: the mass $M(r,t)$ and the spin parameter $a(t,r)$. 
Their forms are the same as in \citet{Janiuk2018}. The code uses dimensionless units, where \mbox{$c=G=M_0=1$}, so the calculated perturbation to the mass of the black hole needs to be divided by the initial mass of the black hole $M_0$ as follows:
\begin{equation}
\begin{array}{c c}
\displaystyle
\xi(t,r) \equiv \frac{M_0 + \Delta M(t) + \delta M(t,r)}{M_0},
& 
\displaystyle
\alpha(t,r) \equiv \frac{J(t,r)}{\xi(t,r)},
\end{array}
\end{equation}
where $\xi$ and $\alpha$ are auxiliary variables introduced here to simplify metric tensor notation. 
The six non-zero components of the self-gravitating Kerr--Schild metric are:
\begin{equation}
\begin{aligned}
g_{tt} &= -1 
  + \frac{2r\xi}{r^2 + \alpha^2 \cos^2\theta}, \\[6pt]
g_{tr} &= \frac{2r\xi}{r^2 + \alpha^2 \cos^2\theta}, \\[6pt]
g_{t\phi} &= -\frac{2\alpha r \xi \sin^2\theta}
                 {r^2 + \alpha^2 \cos^2\theta}, \\[6pt]
g_{rr} &= 1 
  + \frac{2r\xi}{r^2 + \alpha^2 \cos^2\theta}, \\[6pt]
g_{r\phi} &= -\frac{2\alpha r \xi \sin^2\theta}
                 {r^2 + \alpha^2 \cos^2\theta}, \\[6pt]
g_{\phi\phi}
&= \sin^2\theta \left[ \bigl(r^2 + \alpha^2 \cos^2\theta\bigr)
  + \frac{2\alpha^2 r \xi \sin^2\theta}
         {r^2 + \alpha^2 \cos^2\theta} \right].
\end{aligned}
\end{equation}
\subsection{GRMHD scheme}
\label{sec:Numerical_implementation}
\indent We use \texttt{HARM-SELFG}, which is our updated 
version of the general relativistic magnetohydrodynamics code \texttt{HARM} \citep{Gammie2003,Noble2006,2019ApJ...873...12S}. The numerical scheme is conservative and shock-capturing, using the HLL method for calculating fluxes \citep{Harten1983}. The code evolves the system of GRMHD equations representing the conservation of rest-mass, energy-momentum, and induction equations, expressed as:
\begin{equation}
\begin{array}{c c c}
\nabla_\mu\left(\rho u^\mu\right) = 0, & \nabla_\mu \left(T^{\mu\nu}\right) = 0, & \nabla_\mu \left(u^{\mu} b^{\nu} -u^{\nu} b^{\mu}\right)=0,
\end{array}    
\end{equation}
where $u^{\mu}$ is the contravariant four-velocity, $b^{\mu}$ is the contravariant magnetic four-vector, and $\rho$ is the rest-mass density. The numerical scheme uses the MHD stress-energy tensor $T^{\scriptscriptstyle \mu\nu}_{\scriptscriptstyle \mathrm{MHD}}$, which is the sum of the stress-energy tensor for a perfect fluid $T^{\scriptscriptstyle \mu\nu}_{\scriptscriptstyle \mathrm{fluid}}$ and electromagnetic field $T^{\scriptscriptstyle \mu\nu}_{\scriptscriptstyle \mathrm{EM}}$, defined as: 
\begin{equation}
T_{\mathrm{fluid}}^{\mu \nu}=\left(\rho+u+p\right)u^{\mu}u^{\nu}+pg^{\mu \nu},
\end{equation}
\begin{equation}
\label{eq:T_EM}
T_{\mathrm{EM}}^{\mu \nu}=b^{2} u^{\mu}u^{\nu}+\frac{1}{2} b^{2}g^{\mu \nu}-b^{\mu}b^{\nu},
\end{equation}
\begin{equation}
\label{eq:MHD}
T_{\mathrm{MHD}}^{\mu \nu}=T_{\mathrm{fluid}}^{\mu \nu}+T_{\mathrm{EM}}^{\mu \nu}.
\end{equation} 
The initial conditions are specified using the Boyer--Lindquist coordinates \citep{BLcor}, and then transformed to the Kerr--Schild coordinates. The integration of partial differential equations is done in the modified Kerr--Schild coordinates \citep{Noble2006}, with a radial logarithmic grid $x^{[1]}$, where radius $r$ is replaced by \mbox{$r=e^{x^{[1]}}$}. The latitude $\theta$ and azimuthal $\phi$ coordinates remain unchanged. The outer radius of the computational grid is set to \mbox{$1000\,r_{g}$}, with open boundary conditions, allowing free outflow of mass. The boundary conditions in polar and azimuthal directions are reflecting and periodic, respectively.\\
\indent The final time $t_{{f}}$ of all simulations is set to \mbox{$50,000\, t_g$}. The corresponding conversions from code units to CGS units are as follows: \mbox{$r_g  = 4.43 \times 10^{5}\,\mathrm{cm}$} and \mbox{$t_g=1.48 \times 10^{-5}\,\mathrm{s}$}. These units are calculated for \mbox{$M_{\mathrm{BH}}=3\,M_{\odot}$}, and are held constant, in contrast to the size of the event horizon, which increases over time. The computational cost of the model with evolving mass and spin of the black hole in high resolution, such as \texttt{Model-1-NSG} is about \mbox{$\approx900,000$} CPU hours\footnote{The computations were performed on the \href{https://guide.plgrid.pl/en/resources/supercomputers/helios}{Helios supercomputer} at the Academic Computer Centre – Cyfronet AGH, Kraków, Poland.}, whereas the model with self-gravity perturbations is about \mbox{$\approx20\,\%$} more expensive, exceeding one million CPU hours. The scalability is discussed in \citet{JANIUK2025150}. In general, our implementation of self-gravity is an extra cost of about \mbox{$\approx15-25\,\%$} more CPU hours, which we regard as a reasonable expense for a more physically motivated model of collapsar. 

\section{Results}
\begin{figure*}[!t]
    \centering
\begin{subfigure}[b]{0.49\textwidth}
    \centering
    \includegraphics[width=\textwidth, trim=14 9 42 28, clip]{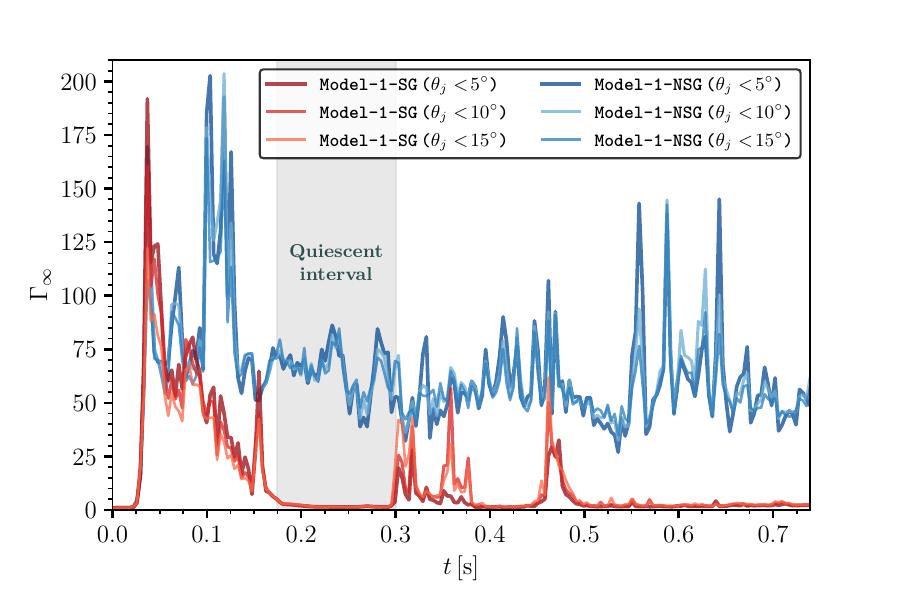}
\end{subfigure}
\begin{subfigure}[b]{0.49\textwidth}
    \centering
    \includegraphics[width=\textwidth, trim=14 9 42 28, clip]{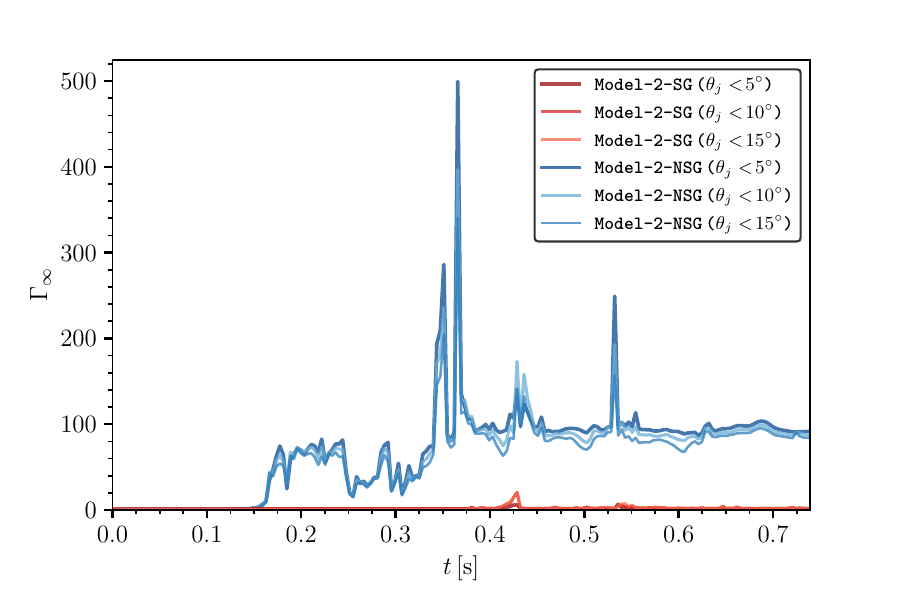}
\end{subfigure}
\caption{Evolution of the terminal Lorentz factor ($\Gamma_{\infty}$) for \texttt{Model-1} (left) and \texttt{Model-2} (right).}
\label{fig:TLF}
\end{figure*}
\subsection{Jet energetics}
\label{sec:Jet_energetics_and_properties}
Astrophysical jets are highly collimated relativistic outflows. The main mechanism responsible for launching Poynting jets is the Blandford--Znajek process \citep{1977MNRAS.179..433B}. In these jets, most of the energy is transported via electromagnetic fields rather than the kinetic energy of particles, and the Maxwell stress can even efficiently collimate the outflows provided the jets at launch are narrow \citep{2011MmSAI..82...95K}. Then, energy is converted into the bulk kinetic energy of particles and emitted as highly energetic radiation \citep{2010MNRAS.402..353L}, observed when the jet axis is pointed towards the observer \citep{Piran2004}. The observed median bulk Lorentz factors of \mbox{$\Gamma\approx300$} \citep{2018A&A...609A.112G} and values in the range \mbox{$\Gamma\approx130-300$} \citep{2024A&A...685A.166R}, indicate that jets associated with GRBs are highly relativistic. In the previous two-dimensional study of self-gravitating collapsars \citep{JaniukSG}, no jet was produced. Below, we describe, for the first time, the effects of self-gravity on the properties of jets launched in our three-dimensional simulations.\\
\indent We measure the jet energetics and variability 
using a parameter defined as:
\begin{equation}
    \mu = \Gamma_{\infty}=-\frac{T^{r}_t}{\rho u^{r}},
    \label{mu_jet}
\end{equation}
where $T^{r}_t$ is the component of the energy‐momentum tensor, representing the radial energy flux. We use the full stress-energy tensor, which contains both the fluid and electromagnetic parts (Eq.~\ref{eq:MHD}). The quantity $\mu$ provides the estimation of the terminal Lorentz factor $\Gamma_{\infty}$, under the assumption that all forms of energy are converted at infinity to the baryon bulk kinetic motion \citep{2003ApJ...596.1104V,2019ApJ...873...12S,2022ApJ...935..176J}.\\
\indent The method for determining $\Gamma_{\infty}$ from our three-dimensional simulations is aimed at finding specific moments in time where jet ejection declines or ceases completely. We measure the terminal Lorentz factor in cones with half-opening angles $\theta_j$ of $5^\circ$, $10^\circ$, and $15^\circ$, extending from the horizon radius $r_h$ to a distance of \mbox{$150\,r_g$}. As a first step, we interpolate the results onto a uniform grid to remove a bias of modified Kerr--Schild coordinates, where most of grid cells are located close to the event horizon. Then, we average the \mbox{$10\,\%$} most energetic elements in the cone. Our method emphasizes contributions from the most energetic elements and is sensitive to physical interruptions in jet ejection to properly probe its variability.\\
\indent The time evolution of terminal Lorentz factors for all models is shown in Fig.~\ref{fig:TLF}. We observe a significant difference between the models with and without self-gravity; however, the initial magnetisation is affecting the whole pattern. First, we note that the choice of the cones half-opening angles $\theta_j$ does not affect the emission patterns; however, for smaller angles, maximum spikes are more energetic, which is related to non-uniform jet energy distribution \citep{2008MNRAS.388..551T,JaniukJames2022,2025A&A...702A..13S}. In \texttt{Model-1-SG} and \texttt{Model-1-NSG}, the jets are launched simultaneously, suggesting that self-gravity does not affect the time of jet formation if it is feasible (as supported by a strong magnetic field). Since the jets originate in the vicinity of the black hole, they are initially dominated by the gravitational influence of the central object. The maximum Lorentz factor in both models is of the same order, and in \texttt{Model-1-SG}, it attains the value of \mbox{$\Gamma_{\infty}\approx190$}, whereas in \texttt{Model-1-NSG}, the maximum value achieves \mbox{$\Gamma_{\infty}\approx200$}.\\ 
\indent In \texttt{Model-2-SG}, the jet ejection does not occur at all, indicating that self-gravity can lead to failed GRBs (in Fig.~\ref{fig:TLF}, the terminal Lorentz factor remains \mbox{$\Gamma_{\infty}\approx 1$} throughout the evolution). We note that in \texttt{Model-2-NSG}, under identical conditions, but without self-gravity, jet ejection occurs, but it is delayed in comparison to \texttt{Model-1-NSG}, which is possibly a consequence of the weaker magnetic field \citep{PhysRevD.110.083014}. In \texttt{Model-1-NSG}, where \mbox{$\beta_0=1$}, jet starts at \mbox{$t\approx0.029\,\mathrm{s}$}, while in \texttt{Model-2-NSG} at \mbox{$t\approx0.158\,\mathrm{s}$}, therefore jet formation time is more than 5 times longer in the model with 10 times less magnetic pressure, \mbox{$\beta_0=10$}. In \texttt{Model-2-NSG}, the terminal Lorentz factor attains at the peak values about \mbox{$\Gamma_{\infty}\approx500$}, and then drops to about \mbox{$\Gamma_{\infty}\approx100$}. A direct comparison of $\Gamma_{\infty}$ values achieved in \texttt{Model-1} and \texttt{Model-2} is challenging due to the resolution difference.\\
\indent For \texttt{Model-1-SG}, and \texttt{Model-1-NSG}, a very sharp initial spike is observed. Then, in the \texttt{SG} case, there is a gradual decline in $\Gamma_{\infty}$, until the jet halts completely. We label this break in the Fig.~\ref{fig:TLF}  
as the Quiescent interval (the threshold is set to \mbox{$\Gamma_{\infty}<5$} for \mbox{$\theta_j=5^{\circ}$}) for \texttt{Model-1-SG}. 
After this break, jet reappears, although it is not as energetic as earlier. At the end of this simulation, the jet in the self-gravitating model ceases, while in the model without self-gravity, it is still present and attains a high Lorentz factor, \mbox{$\Gamma_{\infty}=50-150$}. Our results suggest that in self-gravitating collapsars, a potentially dormant central engine can exist. According to internal shock models \citep{1994ApJ...430L..93R,1997ApJ...490...92K}, it can be responsible for quiescent intervals in the GRB prompt emission. Moreover, we argue that under specific conditions, self-gravity can interrupt the process of jet formation.\\
\afterpage{
\begin{figure*}[t!]
    \begin{subfigure}[b]{0.49\textwidth}
    \centering
    \includegraphics[width=\textwidth, trim=10 9 42 28, clip]{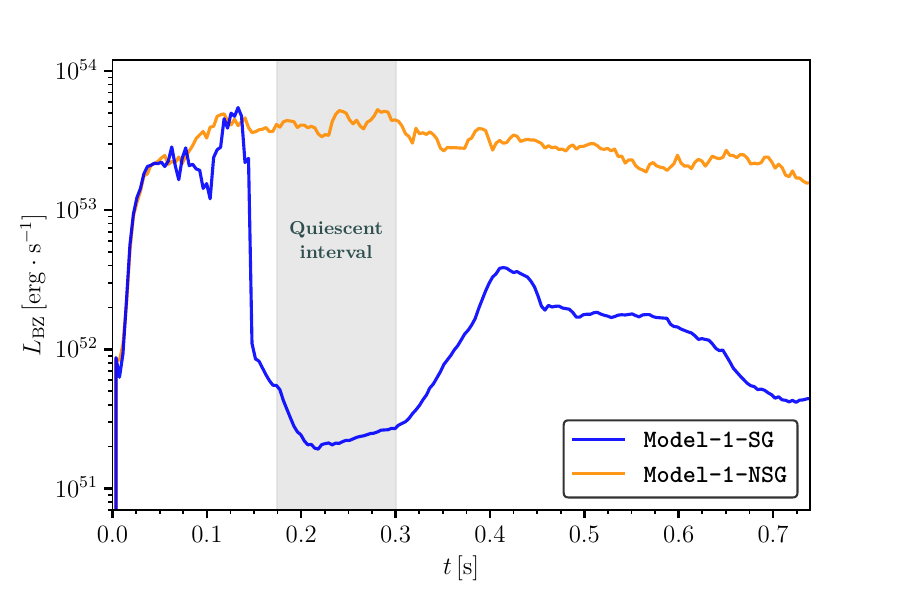}
\end{subfigure}
\begin{subfigure}[b]{0.49\textwidth}
    \centering
    \includegraphics[width=\textwidth, trim=10 9 42 28, clip]{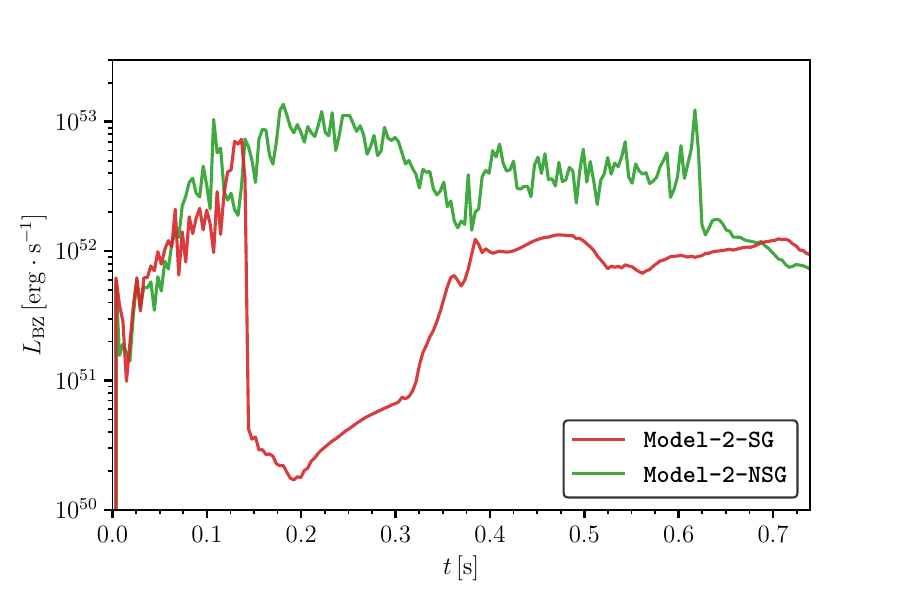}
\end{subfigure}
    \caption{Evolution of Blandford--Znajek luminosity for \texttt{Model-1} (left) and \texttt{Model-2} (right).}
    \label{fig:BZ}
\end{figure*}
}
\indent To obtain the electromagnetic power produced by the system, we compute the Blandford--Znajek luminosity \citep{2004ApJ...611..977M}, defined as:
\begin{equation} 
    L_{\mathrm{BZ}} =-\int_0^{2\pi} \int_0^{\pi}\left(b^{2} u^{r}u_{t}-b^{r}b_t\right)\!\sqrt{-g}\,d\theta \, d\phi,
\end{equation}
and given by the radial energy flux of the electromagnetic stress tensor (Eq.~\ref{eq:T_EM}). In practice, we include only the outgoing electromagnetic energy flux, i.e. we integrate only regions where the radial component of the energy flux is directed outwards. The value is integrated over the event horizon $r_h$, which in our models grows over time due to mass accretion onto the black hole.\\
\indent The evolution of $L_{\rm BZ}$ luminosities for all models
is shown in Fig.~\ref{fig:BZ}. Initially, over the first \mbox{$\sim0.08\,\mathrm{s}$}, in two variants of \texttt{Model-1}, we observe the same jet power. Then, about \mbox{$t\approx0.1\,\mathrm{s}$} in \texttt{Model-1-SG}, there is a small decline. Before the Quiescent Interval, in \texttt{Model-1-SG}, the power rapidly declines, and during the dormant period, it is about two orders of magnitude weaker than in \texttt{Model-1-NSG}. Afterwards, the luminosity grows again, and finally, about \mbox{$t\approx0.4\,\mathrm{s}$}, it diminishes. In contrast, the of energy extraction via the Blandford--Znajek process, in \texttt{Model-1-NSG} is more stable over time, keeping the power about \mbox{$L_{\mathrm{BZ}}\approx (1.57-5.27)\times10^{53}\,\mathrm{erg\,s^{-1}}$}. The stability of \texttt{Model-1-NSG} is further discussed below.\\
\indent\texttt{Model-2-NSG} shows less stable power than \texttt{Model-1-NSG}, with a peak of \mbox{$L_{\mathrm{BZ}}\approx1.36\times10^{53}\,\mathrm{erg\ s^{-1}}$}. In \texttt{Model-2-NSG} at \mbox{$t\approx0.62 \,\mathrm{s}$}, the Blandford--Znajek power declines, despite still large Lorentz factor, \mbox{$\Gamma_{\infty}\sim100$}. This decline is a signature of the gradual jet quenching. In \texttt{Model-2-SG}, we observe the non-zero Blandford--Znajek luminosity, despite the absence of the jet. The pattern is resembling that observed in \texttt{Model-1-SG}, with a deep decrease and finally a secondary increase. However, in \texttt{Model-2-SG}, the outflow cannot be called a jet as it is not collimated.\\
\indent \texttt{Model-2-SG} exhibits the lowest emitted energy $E_j$, about \mbox{$E_j \approx 6.54\times10^{51}\,\mathrm{erg}$}, whereas in \texttt{Model-2-NSG} it is about \mbox{$E_j \approx 3.07\times10^{52}\,\mathrm{erg}$}. The total energy emitted by \texttt{Model-1-SG} is \mbox{$E_j \approx 3.89\times10^{52}\,\mathrm{erg}$}, and for \texttt{Model-1-NSG}, it is about \mbox{$E_j \approx 2.21\times10^{53}\,\mathrm{erg}$}. We note that the ratio of produced energies, \mbox{$\zeta =E_j^{\mathrm{\texttt{SG}}}/E_j^{\mathrm{\texttt{NSG}}}$}, between the models with and without self-gravity is \mbox{$\zeta\approx 0.176$} for \texttt{Model-1}, and for \texttt{Model-2}, it is \mbox{$\zeta\approx 0.213$}. It is worth to note that the ratio of these emitted energies is close to one-fifth, independently of initial $\beta_0$.\\
\indent We suggest that during the Quiescent interval in the central engine, there should be an observable decline in electromagnetic emission from self-gravitating systems. The amount of emitted energy is lower than in the models without self-gravity.
\subsection{Jet opening angle}
\label{sec:Jet_opening_angle}
\begin{figure}[!h]
    \centering
    \includegraphics[width=\linewidth, trim=5 5 5 5, clip]{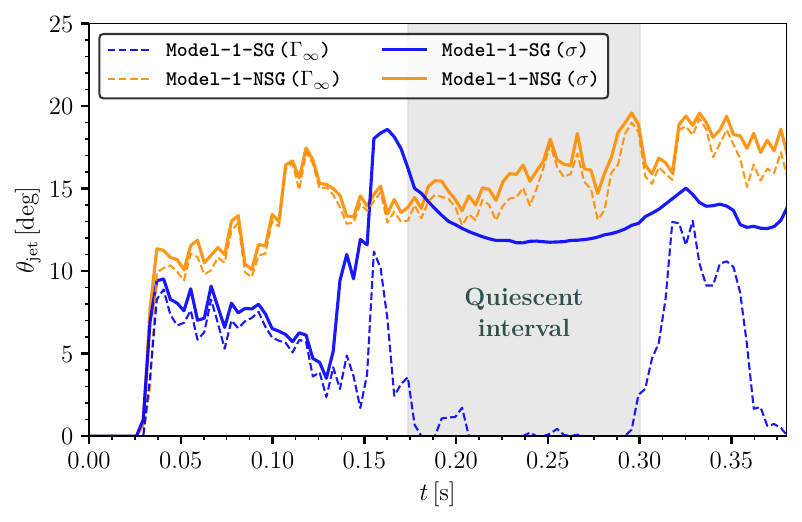}
    \includegraphics[width=\linewidth, trim=5 5 5 5, clip]{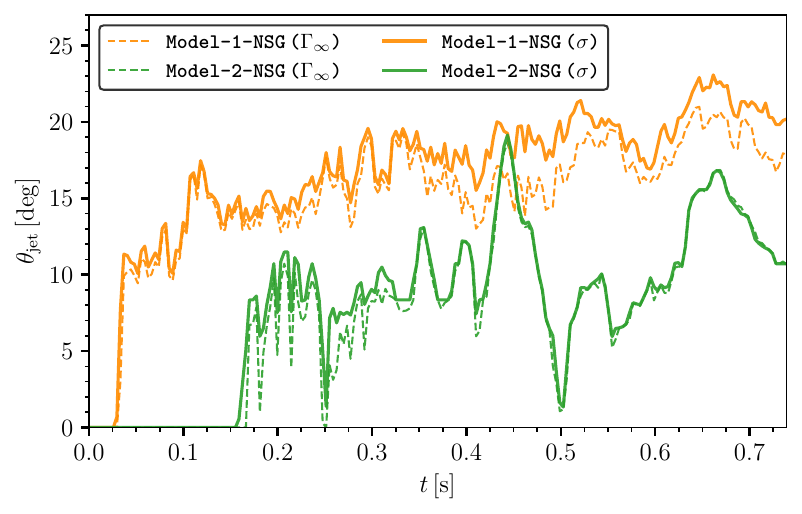}
    \caption{Jet opening angle for \texttt{Model-1} (top), and for \texttt{Model-1-NSG} and \texttt{Model-2-NSG} (bottom), obtained using two methods.}
    \label{fig:OA1}
\end{figure}

Opening angles of GRB jets can be estimated from their afterglow observations by identifying the achromatic break in the multi-wavelength light curve \citep{1999ApJ...519L..17S,2001ApJ...562L..55F}. The analysis of over 100 long GRBs detected by Fermi GBM revealed that the half‐opening angles are well described by a log‐normal distribution, with approximately \mbox{$90\,\%$} of events exhibiting \mbox{$\theta_{\mathrm{jet}}<20^{\circ}$} and the median value of about \mbox{$\theta_{\mathrm{jet}}\approx8^{\circ}$} \citep{Goldstein_2016}. In general, jet collimation is affected by many physical parameters, such as the black hole spin \citep{Hurtado_2024}, magnetic field structure and strength \citep{2006ApJ...651..272F}, disk wind pressure \citep{2016MNRAS.461.2605G}, and the density of the external medium \citep{2013ApJ...764..148L}.\\
\indent To measure the opening angles of jets in our simulations, we use two methods. The first one relies on the magnetization $\sigma$, defined as the ratio of energy stored in the magnetic field to the energy stored in the rest-mass density, given by:
\begin{equation}
    \sigma = \frac{b^2}{\rho}, \quad  b^2 = b^\mu b_\mu.
    \label{sigma_jet}
\end{equation}
This method is based on identifying the highly magnetized regions where \mbox{$\sigma>1$}. In three-dimensional simulations, to avoid inaccuracies resulting from a potential jet-wobbling, we compute opening angles over $N_{\phi}$ cross‐sections. In each cross‐section, we locate the angular distance from the polar northern axis, where \mbox{$\sigma>1$}. The opening angle $\theta_{\mathrm{jet}}$ is then the mean value of all individual half-opening angles measured in all cross-sections. In the second method, similar to the first one, instead of magnetization, we use criterion \mbox{$\Gamma_{\infty} > 5$}. We compute the opening angle at the radial distance of \mbox{$100\, r_g$}. Using two measures allows us to analyze the highly magnetized regions independently of the jet launching process to determine the characteristics of the funnel.\\
\indent In Fig.~\ref{fig:OA1}, the top panel shows the evolution of the opening angle for two configurations of \texttt{Model-1}. First, we note that both methods of measuring opening angle give consistent results in \texttt{Model-1-NSG}. Nevertheless, in \texttt{Model-1-SG}, we observe that shortly before the Quiescent interval, two measurements start to diverge. This behavior is related to the decay and subsequent change of sign of the velocity $u^r$ (at the beginning of accretion from the polar regions), which in turn reverses the sign in Eq.~\ref{mu_jet}, as we consider only outflowing elements. The magnetization arises mostly from the toroidal component $B^{\phi}$ of the magnetic field.
This indicates that during the break in the jet quenching, while $\Gamma_\infty$ is very low,
the highly magnetized funnel, where previously the jet was ejected, still exists; however, the Poynting flux is unable to transport outward efficiently energy. Moreover, the opening angle in the model with self-gravity gradually decreases as it approaches the Quiescent interval, from \mbox{$\theta_{\mathrm{jet}}\approx10^{\circ}$} to \mbox{$\theta_{\mathrm{jet}}\approx4^{\circ}$}. The ratio of the average gas pressure $\langle P\rangle_{\phi}$ at the edge of the jet funnel in both models during the jet launching phase is  \mbox{$\langle P_{\mathrm{\texttt{SG}}}\rangle_{\phi}/\langle P_\mathrm{\texttt{NSG}}\rangle_{\phi} \approx 1.27$} at \mbox{$t = 0.029\,\mathrm{s}$},  increasing to  \mbox{$\langle P_{\mathrm{\texttt{SG}}}\rangle_{\phi}/\langle P_\mathrm{\texttt{NSG}}\rangle_{\phi} \approx 7.5$} at \mbox{$t = 0.125\,\mathrm{s}$}. Then, during the Quiescent interval, we observe the opposite relation. Therefore, we find that due to additional pressure from self-gravitating fluid, the opening angle of the jet funnel is smaller.\\
\indent In Fig.~\ref{fig:OA1}, the bottom panel shows the evolution of opening angles for non-self-gravitating models, 
\texttt{Model-1-NSG} and \texttt{Model-2-NSG}. The jet launching is delayed about five times in \texttt{Model-2}, as shown in Sect.~\ref{sec:Jet_energetics_and_properties}. The chaotic nature of jet is observed in the \texttt{Model-2-NSG}; nevertheless, the opening angle in this case is smaller than in \texttt{Model-1-NSG}, taking the mean value of about \mbox{$\langle \theta_{\mathrm{jet}} \rangle_t \approx 10.2^\circ$}. This is related to the retarded jet ejection and higher inner pressure, when the jet is formed. In \texttt{Model-1-NSG}, we observe the systematic increase of the opening angle over time, from \mbox{$\theta_{\mathrm{jet}}\approx 11 ^\circ$} up to \mbox{$\theta_{\mathrm{jet}}\approx 22 ^\circ$}, with the mean value about  \mbox{$\langle \theta_{\mathrm{jet}} \rangle_t \approx 17.6^\circ$}. This gradual increase is related to the declining spin \citep{Hurtado_2024}, as discussed in below.\\
\indent To sum up, our results show that self-gravity contributes to jet collimation and keeps the opening angles between \mbox{$\theta_{\mathrm{jet}}=4^\circ - 10 ^\circ$}, consistently with observations \citep{Goldstein_2016}.
\begin{figure*}[!t]
\centering
\begin{subfigure}[b]{0.33\textwidth}
\centering
\includegraphics[width=\textwidth, trim=0 0 43 28, clip]{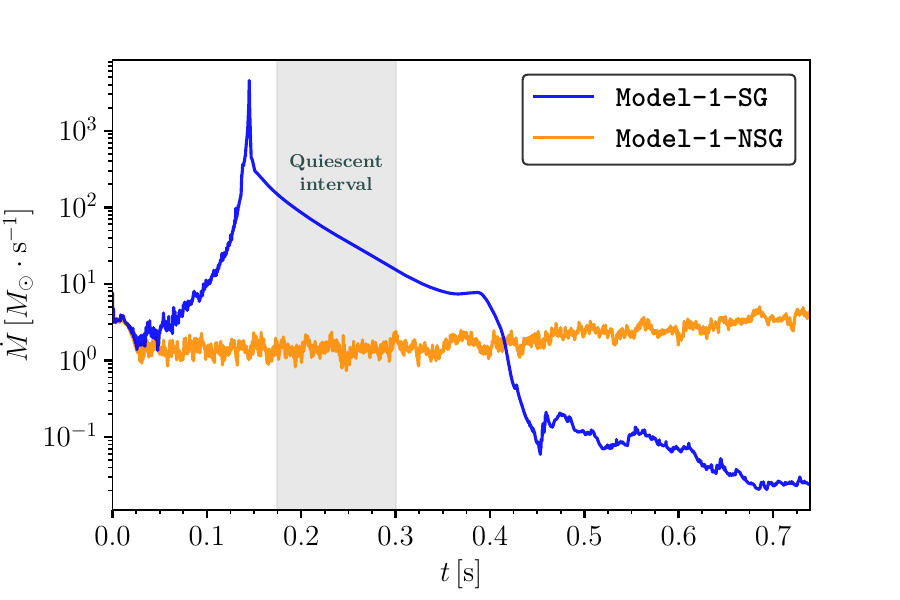}
\end{subfigure}\hfill
\begin{subfigure}[b]{0.33\textwidth}
\centering
\includegraphics[width=\textwidth, trim=0 0 43 28, clip]{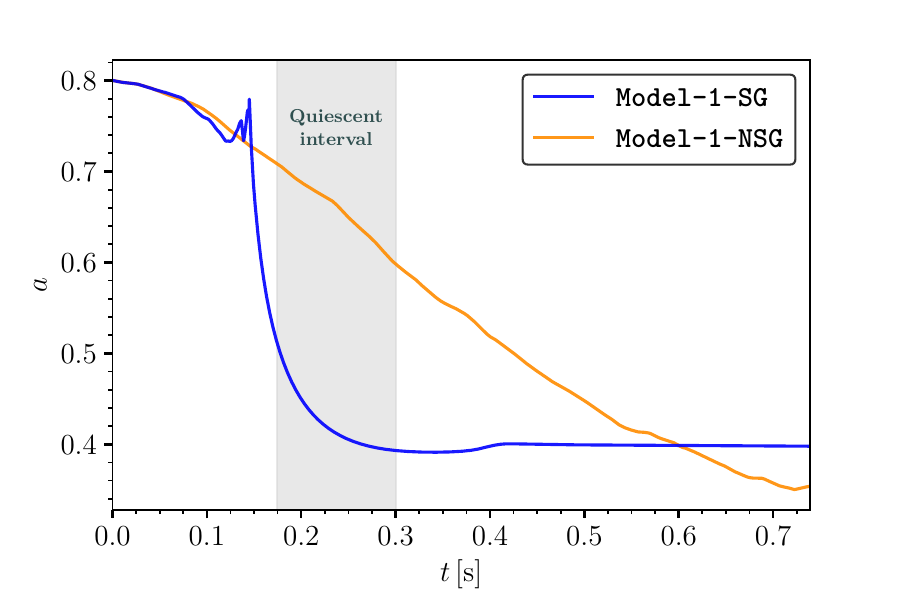}
\end{subfigure}\hfill
\begin{subfigure}[b]{0.33\textwidth}
\centering
\includegraphics[width=\textwidth, trim=0 0 43 28, clip]{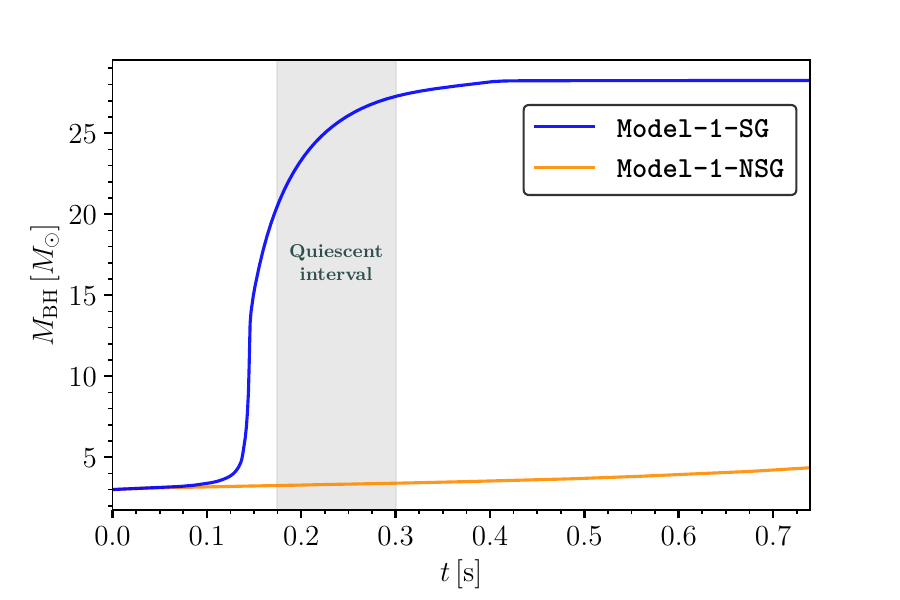}
\end{subfigure}
\begin{subfigure}[b]{0.33\textwidth}
\centering
\includegraphics[width=\textwidth, trim=0 0 43 28, clip]{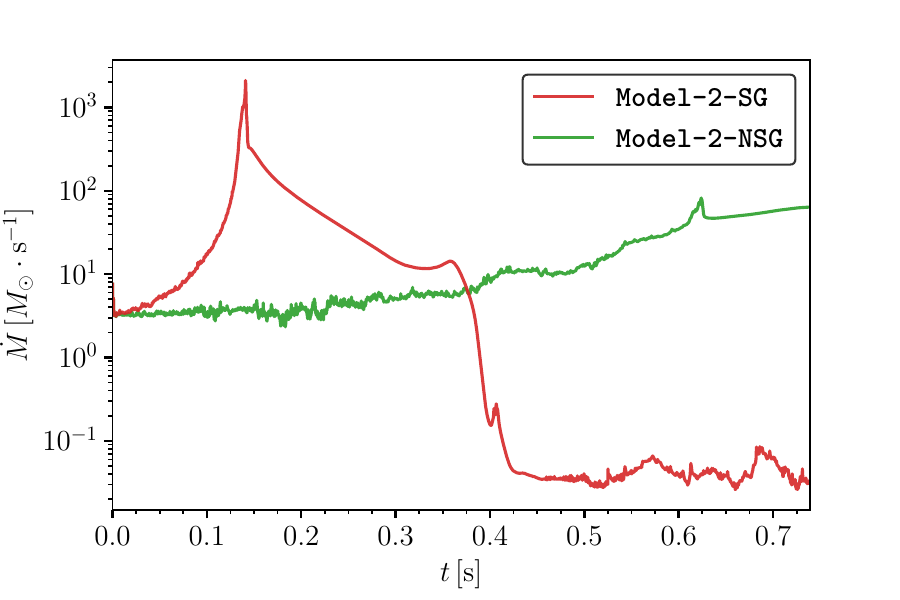}
\end{subfigure}\hfill%
\begin{subfigure}[b]{0.33\textwidth}
\centering
\includegraphics[width=\textwidth, trim=0 0 43 28, clip]{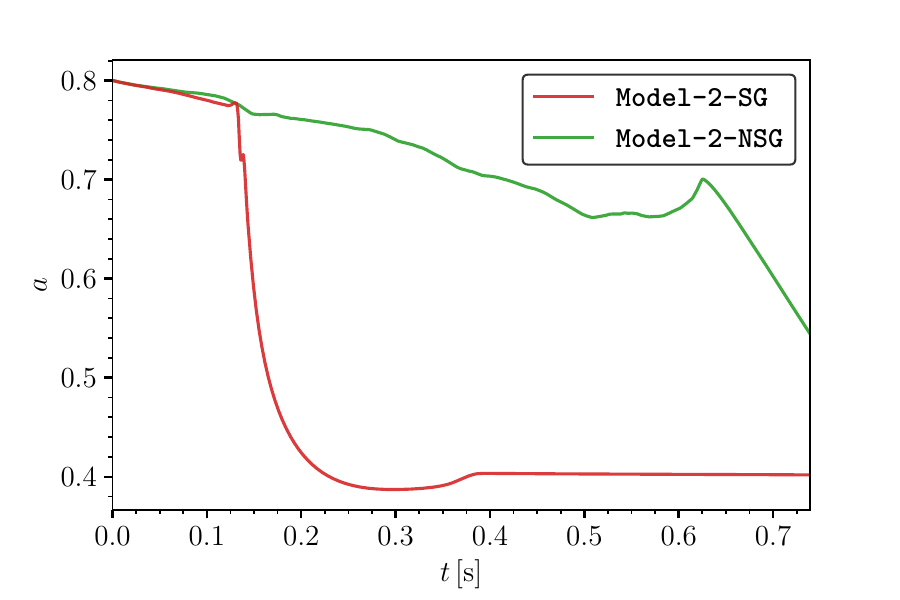}
\end{subfigure}\hfill%
\begin{subfigure}[b]{0.33\textwidth}
\centering
\includegraphics[width=\textwidth, trim=0 0 43 28, clip]{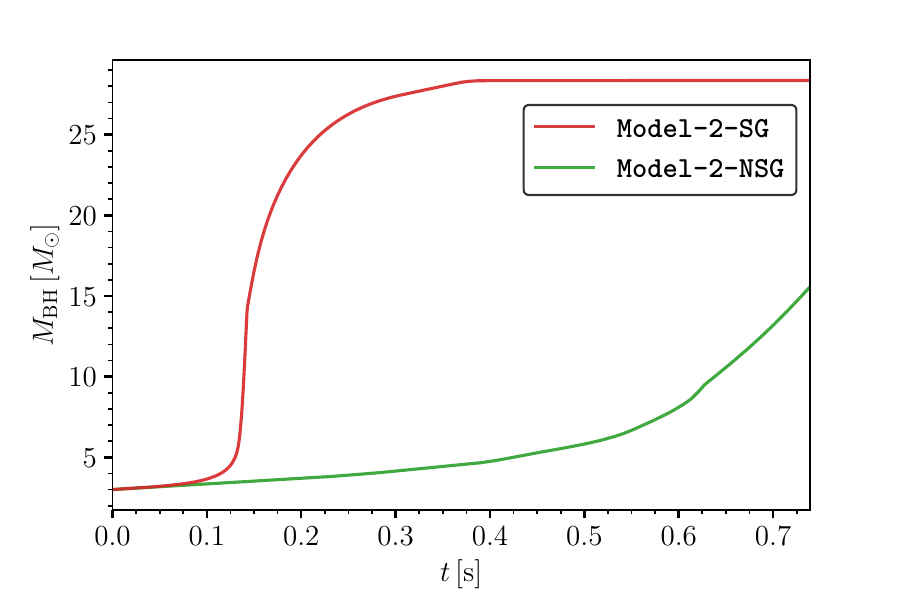}
\end{subfigure}
\caption{Evolution of the black hole spin, mass and accretion rate for \texttt{Model-1} (top) and \texttt{Model-2} (bottom).}
\label{fig:ener.out}
\end{figure*}

\subsection{Black hole mass and spin evolution}
\label{sec:Time-dependent evolution}

Our simulations account for the Kerr metric update due to the change in mass and spin of the black hole, allowing us to track the evolution of these parameters and relate then to jet properties. The mass, spin, and accretion rate onto a black hole are shown in Fig.~\ref{fig:ener.out}, as a function of time.\\
\indent The difference between models with and without self-gravity is clearly visible. The mass accretion rate onto a black hole is significantly larger in models with self-gravity, temporarily exceeding \mbox{$\dot{M}>10^3\, M_{\odot}\, s^{-1}$} in \texttt{Model-1-SG}, while in \texttt{Model-1-NSG} the accretion rate only slowly grows over time, reaching at most \mbox{$\dot{M}\approx8\, M_{\odot}\, s^{-1}$}. The mean value of $\dot{M}$ for \texttt{Model-1-SG} and \texttt{Model-2-SG} in the time interval \mbox{$t <0.4\,\mathrm{s}$} is approximately \mbox{$\dot{M} \approx 60~M_{\odot}\,\mathrm{s^{-1}}$}. In the previous study \citep{JaniukSG}, we observed accretion rates reaching up to \mbox{$\dot{M}\approx500\, M_{\odot}\, s^{-1}$} in the self-gravitating models, which is lower than in the current, three-dimensional study. This is explained by the different initial conditions, and the fact that rapid accretion is preceded by mass accumulation, before the rapid collapse. 
In \texttt{Model-1-SG}, we note that the Quiescent interval is prefaced by the rapid mass accretion, and appears when the accretion rate after the peak starts to decrease. On the contrary, \texttt{Model-1-NSG} and \texttt{Model-2-NSG} exhibit less rapid accretion; nevertheless, the rate is increasing over time, more steeply in \texttt{Model-2-NSG}. In both models, \texttt{Model-2-SG} and \texttt{Model-1-SG}, the accretion rate follows a similar pattern, linked to the perturbations imposed by self-gravity, which produce dramatic effects and cannot be reduced by a strong magnetic field. In models without self-gravity, due to various magnetic pressures, the accretion rate in \texttt{Model-2-NSG} is about one order of magnitude larger than in the more magnetized \texttt{Model-1-NSG}.\\
\indent The spin evolution varies across the models. Initially, in all versions, the evolution proceeds similarly for \texttt{SG} and \texttt{NSG} cases. The spin change is a result of black hole rotational energy extraction, as well as accretion of massive fluid that possesses certain specific angular momentum \citep{2004ApJ...602..312G,refId0,2005ApJ...620...59S}. In \texttt{Model-1-NSG}, the gradual spin decline is related to the Blandford--Znajek process (as outflow of the angular momentum, see Eq.~\ref{eq:spin_angular_momentum}), while in \texttt{Model-1-SG}, the rapid spin decay is not due to extraction, as the Blandford–Znajek luminosity in this model is low (see Fig.~\ref{fig:BZ}), but mostly results from accretion and the increasing mass of the black hole. For self-gravitating models, the final spins for \texttt{Model-1-SG} and \texttt{Model-2-SG} are about \mbox{$a\simeq 0.4$} (in \texttt{Model-1-SG} slightly lower), whereas in the case of the \texttt{Model-1-NSG}, even though most of the mass is not accreted, the spin is lower, \mbox{$a\approx 0.35$}. This result indicates that due to longer and stable jet production, models that do not include self-gravity can extract more rotational energy from the black hole. This trend is visible in both sets of simulations, even though for \texttt{Model-2-NSG} the final spin value was not yet reached. We conclude that self-gravitating collapsars leave black holes with somewhat higher spins, and the equilibrium value is reached earlier than for non-self-gravitating ones, for the same spin at birth.\\
\indent The same effect is visible in the black hole mass evolution. In the right panels of Fig.~\ref{fig:ener.out}, we note that black hole masses evolve much more rapidly in self-gravitating models. The Quiescent interval appears when the mass of the black hole is almost maximal. In \texttt{Model-1-SG} and \texttt{Model-2-SG}, after the evolution, the final mass of the black hole is \mbox{$M_{\mathrm{BH}}\simeq28\,M_{\odot}$}. We note that the accretion rate in \texttt{Model-1-NSG} is relatively low, and after the \mbox{$50,000\, t_g$}, the mass reaches \mbox{$M_{\mathrm{BH}}\approx 5\,M_{\odot}$}, whereas in \texttt{Model-2-NSG}, it is \mbox{$M_{\mathrm{BH}}\approx 16\,M_{\odot}$}, showing an immense effect of larger initial magnetization on the evolution timescale. However, this effect is observed only in the models without self-gravity.
\subsection{Accretion flow structure and MAD state}
\label{sec:Accretion_flow_and_MAD_state}

\begin{figure}[!h]
    \centering
    \includegraphics[width=1\linewidth, trim=80 50 25 15, clip]{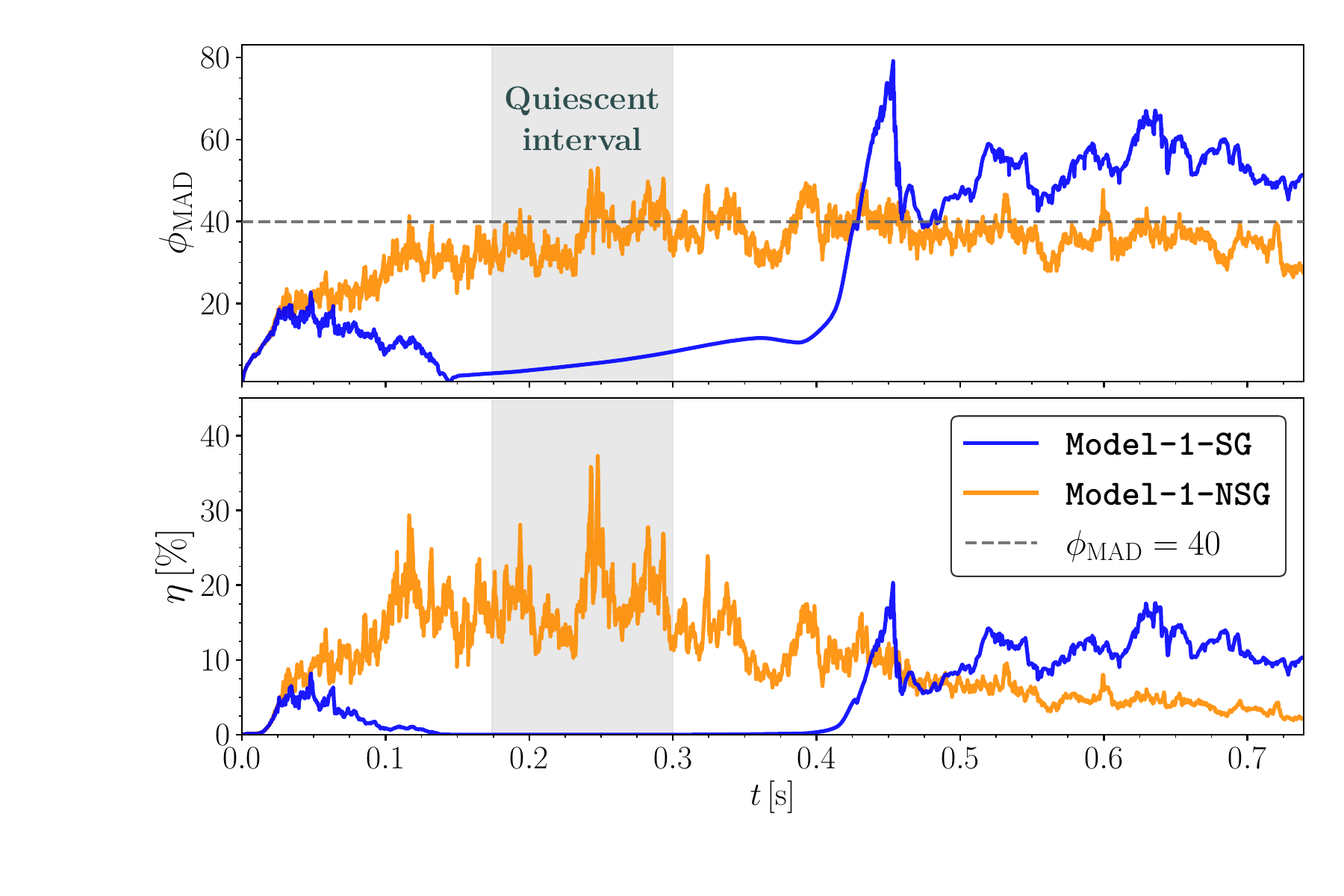}
    \includegraphics[width=1\linewidth, trim=80 50 25 15, clip]{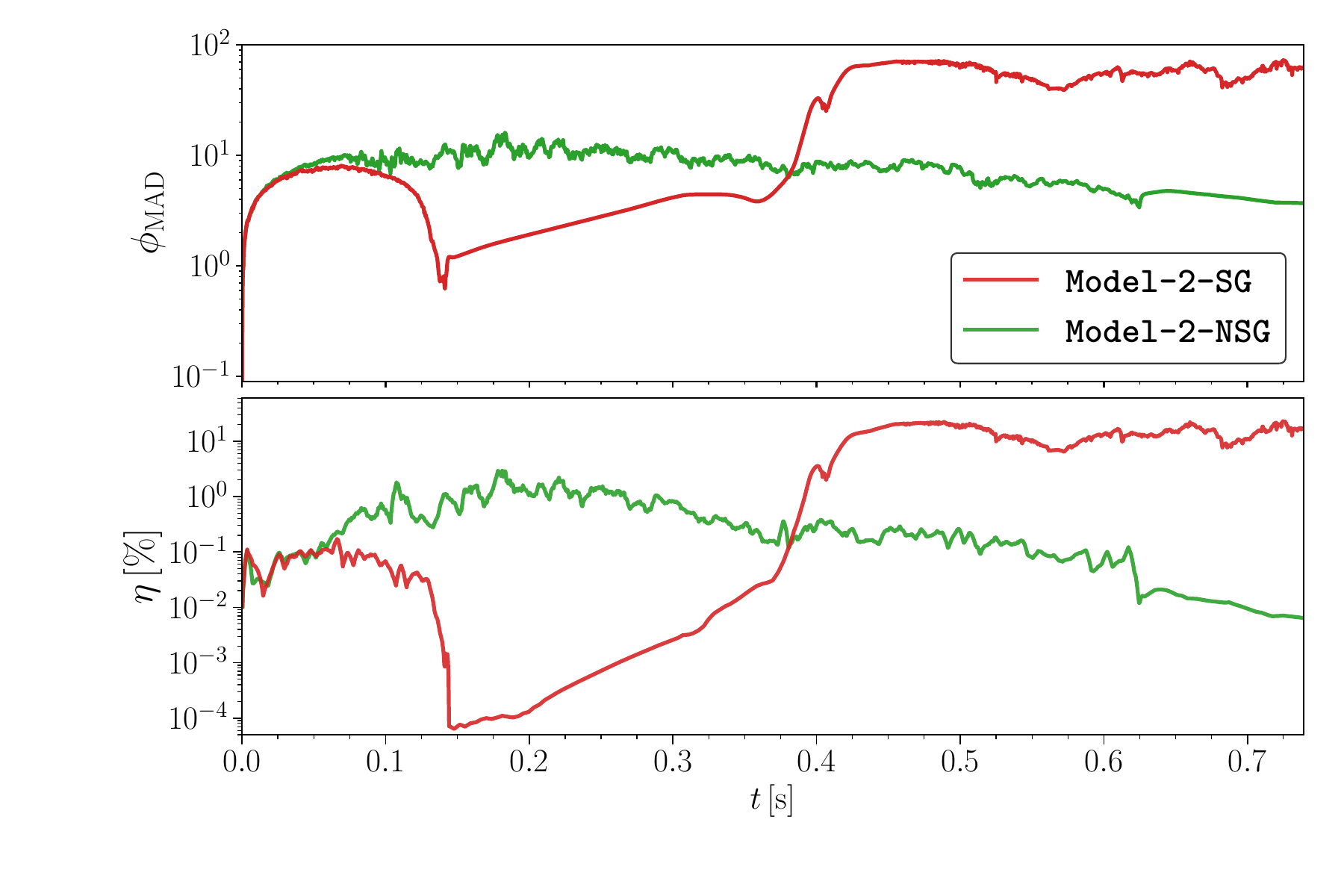}
    \caption{Dimensionless magnetic flux ($\phi_{\rm MAD}$) and emission efficiency ($\eta$) for \texttt{Model-1} and \texttt{Model-2}.}
    \label{fig:MAD1}
\end{figure}

\begin{figure*}[ht]
    \centering
  \includegraphics[width=1\linewidth, trim=85 5 76 28, clip]{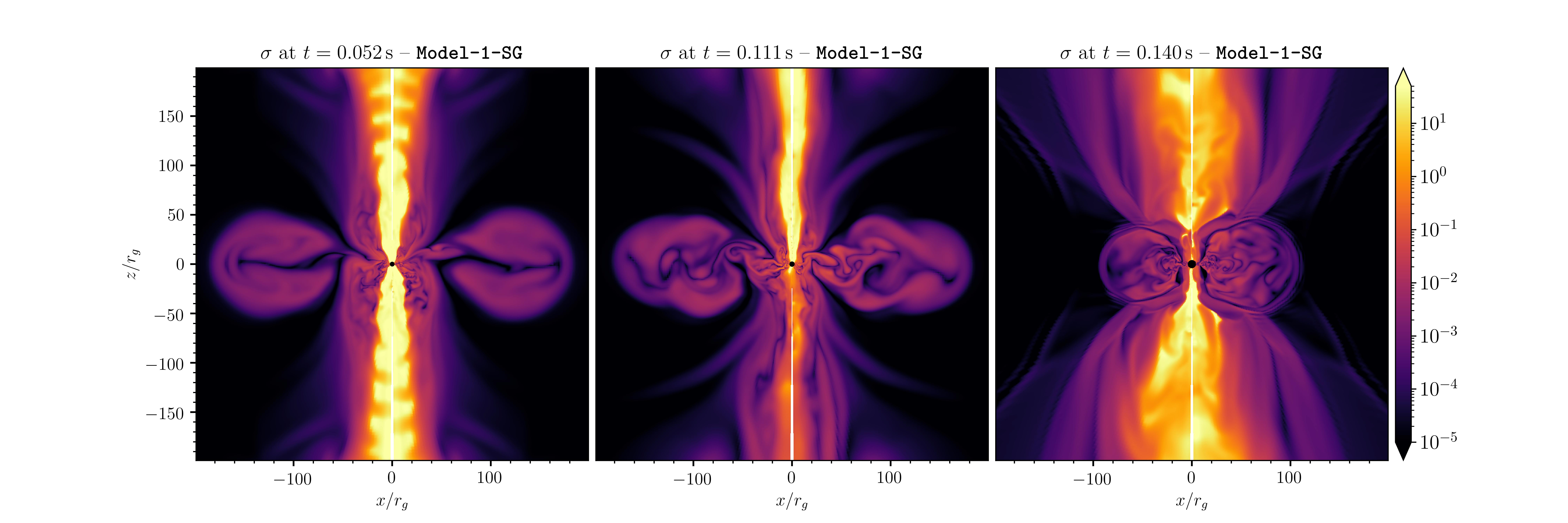}
  \includegraphics[width=1\linewidth, trim=85 5 76 28, clip]{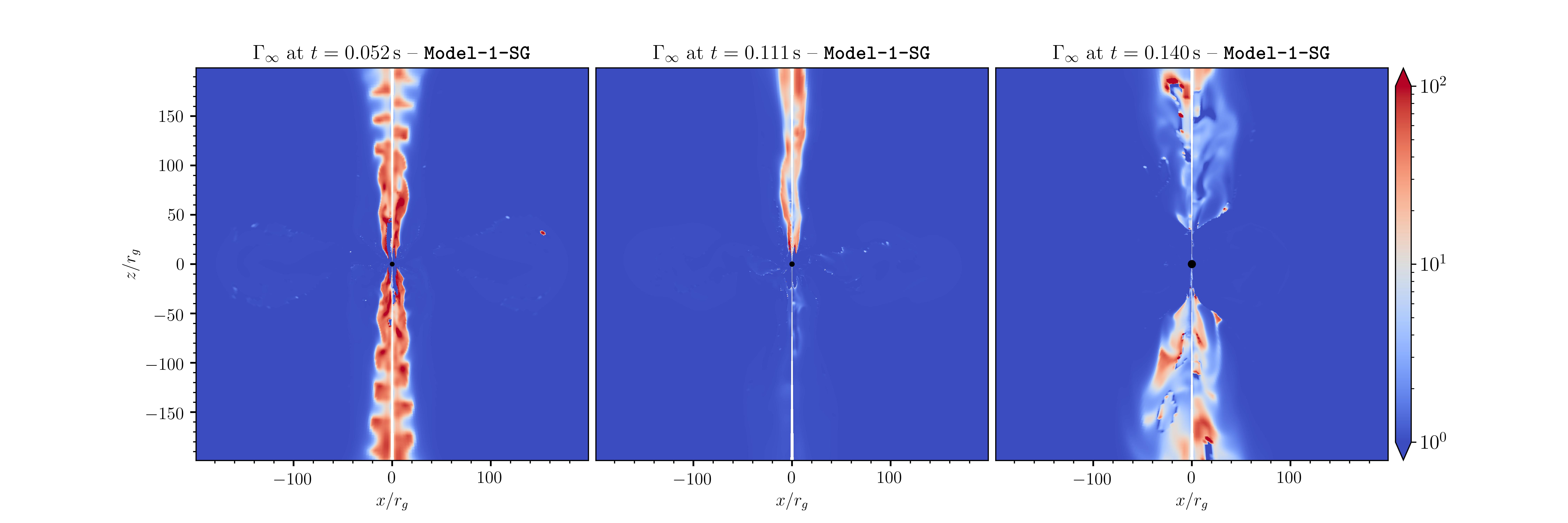}
  \includegraphics[width=1\linewidth, trim=85 5 76 28, clip]{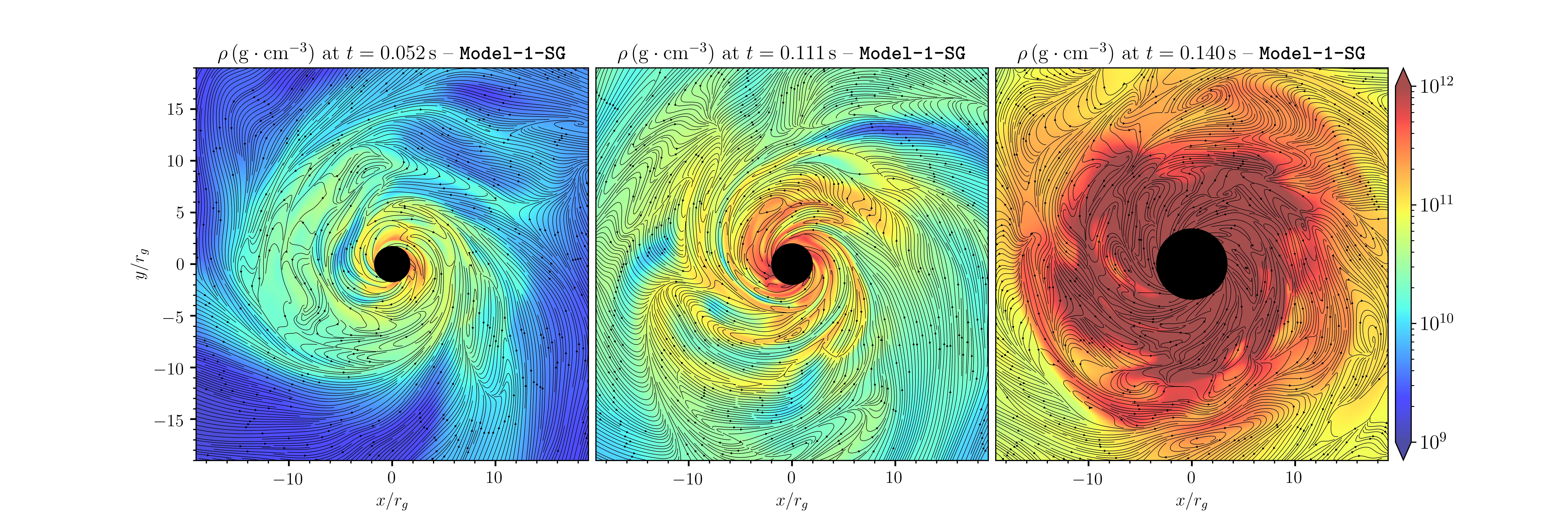}
\caption{Two-dimensional maps of magnetization ($\sigma$) and terminal Lorentz factor ($\Gamma_{\infty}$), and equatorial-plane map of rest-mass density ($\rho$) with overlaid magnetic field lines for \texttt{Model-1-SG}. Columns show $t = 0.074$, $0.111$, and $0.140\,\mathrm{s}$.}

    \label{fig:sigma_mu_rho}
\end{figure*}

\begin{figure*}[t]
\includegraphics[width=\linewidth, trim=105 30 80 50, clip]{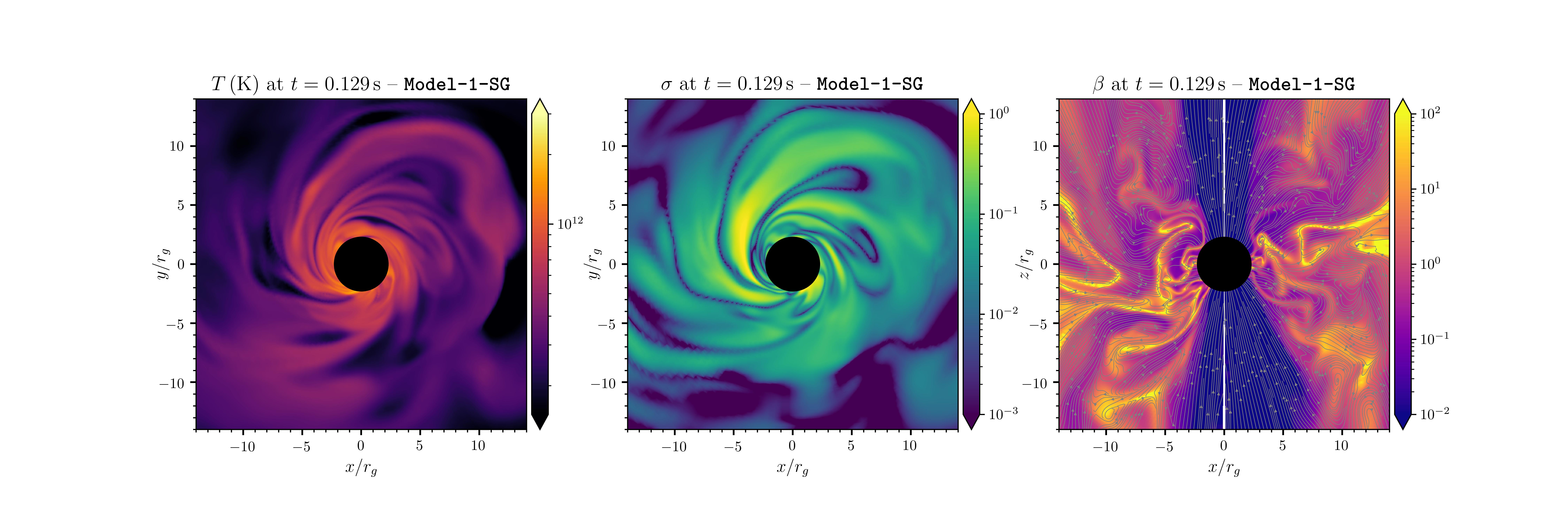}
\includegraphics[width=\linewidth, trim=105 30 80 50, clip]{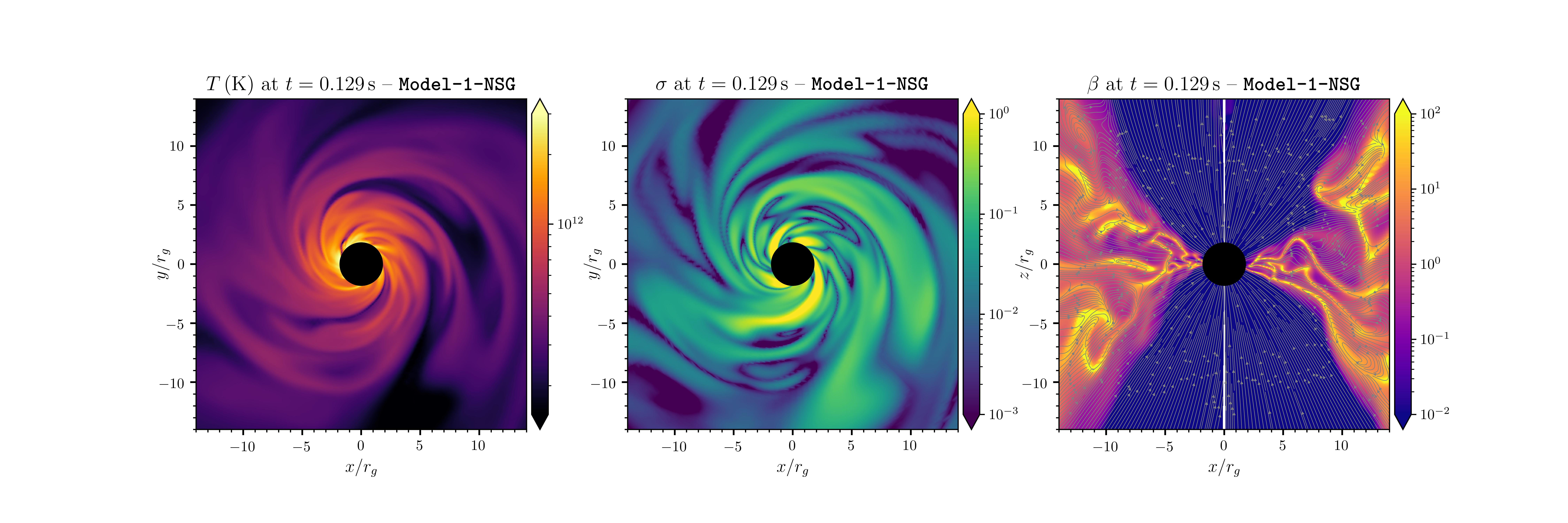}
        \caption{Two-dimensional maps of  temperature ($T$), magnetisation ($\sigma$) and plasma beta ($\beta$) of \texttt{Model-1-SG} (top row) and \texttt{Model-1-NSG} (bottom row). All snapshots correspond to \mbox{$t = 0.129\, \mathrm{s}$}.}
    \label{fig:fig2}
\end{figure*}

Magnetized accretion onto the black hole is mostly driven by the magnetorotational instability (MRI, \cite{1991ApJ...376..214B}). The magnetically arrested disk (MAD) is the model proposed to explain the properties of powerful jets. Early analytic works (e.g. \citealt{1974Ap&SS..28...45B,1976Ap&SS..42..401B}), pointed out that a sufficiently strong accumulation of magnetic flux near the event horizon can efficiently suppress accretion. It was noted in GRMHD simulations \citep{2003PASJ...55L..69N,2003ApJ...592..767P,2003ApJ...592.1042I}, that the effect can become so strong that it almost chokes the accretion flow. In the MAD state, the accumulation of strong magnetic flux near the black hole can suppress the MRI, leading to the dominance of magnetic Rayleigh--Taylor (RTI) instabilities, also known as interchange instabilities \citep{2018MNRAS.478.1837M,2019ApJ...874..168W}.\\ 
\indent The mass inflow strongly interacts with the magnetic barrier at the magnetospheric radius $r_{m}$, where the force of gravity is equal to the magnetic force. The effect can be conveniently quantified under the assumption that all magnetic flux below $r_m$ reaches the black hole horizon \citep{2012MNRAS.423.3083M}. However, recent three-dimensional simulations have suggested that in strongly magnetized flows, accretion can be controlled by chaotic magnetic disconnections and local flux eruptions, not by a global MAD-type barrier \citep{2024A&A...692A..37N}.\\
\indent The dimensionless magnetic flux $\phi_{\rm MAD}$ measures the mass-loading of the magnetic field lines and is given by:
\begin{equation}
\phi_{\rm MAD} \;=
\frac{
  \displaystyle\int\!\!\int
  \bigl|\,B^r\bigr|\,
  \sqrt{-g}\;
  d\theta\,d\phi
}{
  \sqrt{\dot{M}\,r_g(t)^2\,c}
}
\end{equation}
where $B^r$ is the radial component of the magnetic field measured at the horizon, and $\dot{M}$ is the mass accretion rate onto the black hole. In this formula, $r_g$ is time-dependent to provide proper length scaling, since the mass of the black hole changes over time. Typically, when $\phi_{\rm MAD}$ exceeds a critical value (about \mbox{$\phi_{\mathrm{MAD}}\approx 50$} in Gaussian units, or about \mbox{$\phi_{\rm MAD}\approx 15$} in Lorentz--Heaviside units; the former are used in this study), the magnetic flux near the horizon saturates and suppresses the mass inflow, causing the disk to enter the MAD state \citep{2003PASJ...55L..69N}. In consequence, a force-free magnetosphere forms and the Blandford–Znajek process is activated, so launching of relativistic jets is highly efficient, as demonstrated by \citet{Tchekhovskoy2011}. To estimate the jet production efficiency $\eta$, we use the following parameter:
\begin{equation}
\eta = \frac{L_{\mathrm{BZ}}}{\dot{M}c^2}.
\end{equation}
\indent The quantities $ \phi_{\rm MAD}$ and $\eta$ for \texttt{Model-1} and \texttt{Model-2} are shown as a function of time, in Fig.~\ref{fig:MAD1}.
Initially for \texttt{Model-1}, we observe that $\phi_{\rm MAD}$ evolves similarly in both models, and in \texttt{SG} and \texttt{NSG} cases, reaching values up to \mbox{$\phi_{\rm MAD}\approx20$}. Subsequently, the models start to diverge. In the model with self-gravity, we observe a gradual decline, whereas in the model without self-gravity, this value saturates temporarily at \mbox{$\phi_{\rm MAD}\approx 40$}, and then slowly decreases over time to \mbox{$\phi_{\rm MAD}\approx 30$}, still kept within the MAD state. In \texttt{Model-1-NSG}, the jet efficiency is relatively high, reaching up to \mbox{$\eta\approx 37\,\%$} (with the black hole spin at that moment \mbox{$a\approx0.65$}), and then slowly declines to \mbox{$\eta\lesssim 5\,\%$}. Even though the $ \phi_{\rm MAD}$ is relatively stable, the efficiency is decreasing due to reduced black hole spin \citep{Tchekhovskoy2011,Lowell_2024}. In \texttt{Model-1-SG}, we note that self-gravity prevents the increase of $\phi_{\rm MAD}$, and reduces its value to about \mbox{$\phi_{\rm MAD}\lesssim 1$}, before the peak in accretion rate. This has an impact on the jet production efficiency, which decreases from \mbox{$\eta\approx 8\,\%$} to \mbox{$\eta\ll 1\,\%$}, before the Quiescent interval. This suggests that in the self-gravitating collapsars, it is much harder to produce highly efficient jets. Self-gravity acts as the main factor responsible for the increasing mass accretion rate; hence, the magnetic field accumulation in the vicinity of the black hole is not able to choke the inflow effectively. The transition after the Quiescent interval, when $\phi_{\rm MAD}$ starts to grow rapidly up to \mbox{$\phi_{\rm MAD}> 50$}, is related to the fact that most of the mass from the system is already accreted, and the numerical floor becomes important. The magnetic flux $\Phi_{\mathrm{BH}}$ on the event horizon in \texttt{Model-1-SG} and \texttt{Model-2-SG} gradually decreases. In \texttt{Model-1-SG}, when the black hole horizon reaches almost the final size at \mbox{$t = 0.4\,\mathrm{s}$}, the flux drops from \mbox{$\Phi_{\mathrm{BH}} = 9.3\times10^{29}\,\mathrm{G\,cm^2}$} to \mbox{$\Phi_{\mathrm{BH}} = 2.6\times10^{29}\,\mathrm{G\,cm^2}$} at \mbox{$t = 0.7\,\mathrm{s}$}. This effect is associated with magnetic reconnection, a process of slowly decaying magnetosphere known as black hole “balding” \citep{2021PhRvL.127e5101B,2024ApJ...968L..10S}.
\\
\indent In \texttt{Model-2}, we observe a similar evolution, as in \texttt{Model-1}. However, the jet efficiency and $\phi_{\rm MAD}$ for \texttt{Model-2-NSG} are not as high as in \texttt{Model-1-NSG}. The dimensionless magnetic flux is about \mbox{$\phi_{\rm MAD}\approx 10$}, and the efficiency oscillates about \mbox{$\eta\lesssim 3\,\%$}. In contrast to \texttt{Model-1}, in the self-gravity scenario of \texttt{Model-2}, the model is not able to achieve the MAD state. We conclude that, due to self-gravity, a higher accretion rate leads to mass accumulation near the black hole, which in turn suppresses jet formation. The Quiescent interval in the models that form a jet can have observational consequences for the jet intermittency.\\
\indent In Fig.~\ref{fig:sigma_mu_rho}, we present poloidal slices of magnetization $\sigma$, terminal Lorentz factor $\Gamma_{\infty}$, and rest-mass density $\rho$ overlaid with magnetic field streamlines taken at the equatorial plane of \texttt{Model-1-SG}. We selected three representative moments in time: (i) when the jet ejection is stable (at \mbox{$t = 0.052\,\mathrm{s}$}, with \mbox{$a\approx0.79$} and 
\mbox{$M_{\mathrm{BH}}\approx3.13\,M_{\odot}$}); (ii) when the ejection becomes unstable due to episodic mass accretion (at \mbox{$t = 0.111\,\mathrm{s}$}, with \mbox{$a\approx0.75$} and 
\mbox{$M_{\mathrm{BH}}\approx3.5\,M_{\odot}$}); and (iii) just before the peak accretion rate, when the jet is completely quenched (at \mbox{$t = 0.140\,\mathrm{s}$}, with \mbox{$a\approx0.74$} and \mbox{$M_{\mathrm{BH}}\approx6.1\,M_{\odot}$}). During phase (i), we observe that magnetization is constant along the jet funnel, and the jet ejection is stable, as is shown in the terminal Lorentz factor map. In the density distribution, we do not observe any large mass accumulation. Subsequently, in phase (ii), the launching becomes unstable. We observe that in the southern hemisphere, the jet launching is suppressed, and ejection takes place only in the northern hemisphere. As seen in the $L_{\mathrm{BZ}}$ in Fig.~\ref{fig:BZ} at about \mbox{$t\approx0.1\,\mathrm{s}$}, as a steep decline in electromagnetic power is a temporary effect lasting about \mbox{$0.01\,\mathrm{s}$}. We note that there is no inherent asymmetry toward the northern hemisphere. This effect is completely random and possibly arises due to the initial perturbation in the internal energy, leading to temporarily greater mass accumulation on one or the other side of the jet. The interesting outcome is that in \texttt{Model-1-SG}, two observers located on the opposite sides of the jet axis would observe different prompt emissions. Finally, during (iii), just before the highest accretion rate peak, we observe that up to \mbox{$50\,r_g$} the jet is quenched in both hemispheres. On the equatorial plane, the mass density starts to exceed \mbox{$\rho\sim10^{12}\,\mathrm{g}\,\mathrm{cm}^{-3}$}. It is worth noting that, when rapid accretion begins, the magnetic field streamlines start to be more chaotic. The Quiescent interval begins here. This moment is also related to the fastest growth of the size of the event horizon.\\
\indent In Fig.~\ref{fig:fig2}, we show two-dimensional maps of temperature $T$, magnetisation $\sigma$ and plasma beta $\beta$ for \texttt{Model-1-SG} (top row) and \texttt{Model-1-NSG} (bottom row). In the distribution of temperature, we observe that accretion is more chaotic in the model with self-gravity, and temperature achieves slightly lower values, which is related to a larger density of infalling matter. The maps of magnetization $\sigma$ show that in \texttt{Model-1-SG}, there are more regions with a low value of magnetization (\mbox{$\sigma \lesssim10^{-3}$}), and magnetization in the vicinity of the black hole is greater in \texttt{Model-1-NSG}. The right panels show the distribution of plasma-$\beta$ in the poloidal plane co-aligned with the jet axis. The most noticeable effect shown in these figures was discussed in Sect.~\ref{sec:Jet_opening_angle}, and this is the jet opening angle. In \texttt{Model-1-NSG}, the opening angle is significantly larger, as in Fig.~\ref{fig:OA1}. Moreover, the streams of infalling matter with higher values of $\beta$ are compressed towards the jet axis, and they are more densely packed around the event horizon of the black hole.\\
\indent To sum up, in our self-gravitating collapsars, the main observable effect is the appearance of the Quiescent interval, related to the mass accumulation and choking of the jet funnel, which results in the jet quenching. Moreover, in the models with self-gravity, it is much harder to maintain the stable MAD state.

\section{Discussion}
\label{sec:Discussion}
In this study, we present first three-dimensional simulations of jet formation and propagation in collapsing self-gravitating stars, extending our prior studies of self-gravity effects in collapsing stars \citep{JaniukSG}. We chose initial conditions to simulate the core of a collapsing massive star endowed with low angular momentum. We found that such a self-gravitating star can produce relativistic jets, if supplied with a strong magnetic field with pressure equal to at least \mbox{$10\,\%$} of the gas pressure at the core interior. The initial fast rotation of the black hole is slowing down during the collapse, but is not critical to sustain a Blandford--Znajek powered jet. Our implementation of self-gravity is an intermediate step between usual GRMHD simulations of collapsars \citep{2022ApJ...933L...9G,2022MNRAS.510.4962G,Gottlieb_2024} and future three-dimensional simulations based on the BSSN formulation. Recent axisymmetric dynamical-spacetime GRMHD simulations have demonstrated the important role of dynamical metric evolution in collapsars \citep{PhysRevD.109.043051,2025PhRvD.111l3017S}.\\
\indent We studied models in the presence and absence of self-gravity under identical initial conditions to obtain a direct probe of jet properties resulting from self-gravitating collapsars. We used here the transonic configuration of a spherical mass distribution around a newly formed black hole, similar to the previously discussed in \citet{Janiuk2018,Dominika2021,JaniukSG}. This distribution simplifies the problem of collapsar initial configuration, that would result from a pre-supernova star simulation and stellar evolution models. Notably, smooth initial density profile is used, to avoid imposing density inhomogeneities which appear because of chemical composition changes in the pre-supernova phase \citep{2000ApJ...528..368H,2005ApJ...626..350H}. These initial inhomogeneities might have an impact on the \texttt{SG} models, because of their interplay with potential hydrodynamical instabilities and disk fragmentation. These effects were earlier discussed in \citep{JaniukSG} and are further explored in our follow-up paper \citep{Plonka2025}.\\
\indent The role of magnetic fields in collapsing stars can be important to halt accretion and limit the black hole spin-up, as discussed previously in \citet{Dominika2021}. In those models, however, only weakly magnetized collapsars and/or slowly rotating black holes were studied. Here we focus on magnetisations high enough to launch bi-polar jets at the cost of the rotation of a spinning black hole. We find the existence of powerful jets in non-self-gravitating collapsars, while the effects of self-gravity are in competition with the gas magnetisation, and have an impact on the MAD state accretion, as already suggested in \citet{JaniukSG}.\\
\indent We show here that self-gravity can lead to dormant periods in the activity of the central engine. We argue that this occurs due to the mass accumulation horizon region and rapid accretion onto the black hole. We observe that in the model without self-gravity, the MAD state remains stable over time, whereas in the model with self-gravity, this state is disrupted. In the self-gravitating case, therefore, we obtain a gradual decay of the jet efficiency, which precedes the Quiescent interval. We observe that the opening angle of the jet of self-gravity is significantly smaller, and it gradually decreases until the jet is completely quenched. In particular, the case simulated by \texttt{Model-1-NSG} more efficiently extracts rotational energy from the black hole, resulting in a smaller value of the final black hole spin. We also speculate that when the initial conditions are not able to produce a jet and allow its breakout quickly enough, it potentially can lead to a failed GRB, like in \texttt{Model-2-SG}. This prediction poses a question whether the initial conditions and their parameter space used in standard simulation without including self-gravity can be used to uniquely constrain the favored scenario for a successful GRB. Our results imply a narrower range of initial conditions describing the pre-collapse star, such as the magnetic field strength, stellar rotation, and black hole spin, that would be capable of producing relativistic jets in collapsars.\\
\indent The prompt emission in gamma-ray bursts exhibits variability on three characteristic time-scales \citep{2002MNRAS.331...40N}. The shortest timescales are related to variability in the central engine. 
The intermediate time-scale is a plateau with no activity, and the longest time-scale corresponds to the timescale of the burst at jet breakout. The origin of intermediate quiescent times remains unclear \citep{2001MNRAS.320L..25R,2002MNRAS.331...40N}. Potentially, some of them can be explained by jet-wobbling motion as it was shown in \citet{2022ApJ...933L...9G}. The longest time scales are related to the entire time of the engine activity. We argue that self-gravitating collapsars can provide a complementary explanation for intermediate time-scales and the appearance of quiescent times. Potentially, the combination of jet wobbling and quiescent intervals can explain a wide range of long gamma-ray burst activity patterns.

\section{Conclusions}
\label{sec:Conclusions}
The main findings of this study are 
as follows:
 \begin{itemize}
     \item Including self-gravity in GRMHD may lead to temporary jet suppression, which could potentially explain some of the quiescent periods observed in the prompt emission of long gamma-ray bursts.
     \item The Quisecent interval observed in \texttt{Model-1-SG} is caused by the accumulation of mass, and then rapid accretion onto the black hole.
     \item There is no substantial difference in the jet launching time between models with and without self-gravity, if such ejection is feasible in both models.
     \item The models in the absence of self-gravity, due to more stable jet ejection, can extract more rotational via Blandford--Znajek process, leaving black holes with lower spin values. 
     \item The presence of self-gravity is an additional factor that decreases the jet opening angle, which is related to the effect of the extra pressure exerted on the jet funnel. 
     \item In the models including self-gravity, it is much harder to achieve a stable magnetically arrested disk state. Therefore, the jet efficiency is significantly reduced in the models with self-gravity.
     \item In weaker magnetized case (\texttt{Model-2-SG}), self-gravity disrupts the jet formation, resulting in a failed GRB with no jet ejection. Under the same initial conditions but without self-gravity, jet launching does occur.
     \item The observation that self-gravity can lead to a failed GRB, suggests that the parameter space of initial conditions capable of producing successful jets is narrower.
     \item The direct comparison between \texttt{Model-1-NSG} and \texttt{Model-2-NSG} shows that the jet launching time is significantly delayed (about 5 times) in the model with weaker magnetization.
     \item Overall, self-gravity reduces the evolutionary timescale of the system, causing the final black hole’s spin and mass to evolve more rapidly. This is consistent with the conclusions from two-dimensional study \citep{JaniukSG}.
 \end{itemize}

\begin{acknowledgements}
We would like to thank Krzysztof Nalewajko, Gerardo Urrutia Sanchez, and Ore Gottlieb for valuable discussions and insightful comments that helped improve this study. This work has been partially supported by grant No. 2023/50/A/ST9/00527 from the Polish National Science Center.
We gratefully acknowledge Polish high-performance computing infrastructure PLGrid (HPC Center: ACK Cyfronet AGH) for providing computer facilities and support within computational grant no. PLG/2025/018232
\end{acknowledgements}

\bibliographystyle{aa}
\bibliography{aselfg}

\end{document}